\documentclass[aps,prd,superscriptaddress,nofootinbib,preprintnumbers,12pt]{revtex4}
\usepackage{amsmath}
\usepackage{graphicx}
\usepackage{dcolumn}
\usepackage{bm}
\usepackage{amssymb}
\usepackage{latexsym}

\newcommand{\be}{\begin{equation}}
\newcommand{\ee}{\end{equation}}
\newcommand{\bq}{\begin{eqnarray}}
\newcommand{\eq}{\end{eqnarray}}
\begin{document}

\title{Features of holographic dark energy under the combined cosmological
constraints}
\author{Yin-Zhe Ma}
\email{yzm20@cam.ac.uk}
\affiliation{Kavli Institute for Theoretical Physics China, Institute of Theoretical
Physics, Chinese Academy of Sciences, P.O.Box 2735, Beijing 100080, China}
\affiliation{Institute of Astronomy, University of Cambridge, Madingley Road, Cambridge,
CB3 0HA, U.K.}
\author{Yan Gong}
\email{gongyan@bao.ac.cn}
\affiliation{National Astronomical Observatories, Chinese Academy of Sciences, Beijing
100012, China}
\author{Xuelei Chen}
\email{xuelei@cosmology.bao.ac.cn}
\affiliation{National Astronomical Observatories, Chinese Academy of Sciences, Beijing
100012, China}
\affiliation{Kavli Institute for Theoretical Physics China, Institute of Theoretical
Physics, Chinese Academy of Sciences, P.O.Box 2735, Beijing 100080, China}

\begin{abstract}
We investigate the observational signatures of the holographic dark energy
models in this paper, including both the original model and a model with an
interaction term between the dark energy and dark matter. We first delineate
the dynamical behavior of such models, especially whether they would have
``Big Rip'' for different parameters, then we use several recent
observations, including 182 high-quality type Ia supernovae data observed
with the Hubble Space Telescope, the SNLS and ESSENCE surveys, 42 latest
Chandra X-ray cluster gas mass fraction, 27 high-redshift gamma-ray burst
samples, the baryon acoustic oscillation measurement from the Sloan Digital
Sky Survey, and the CMB shift parameter from WMAP three years result to give
more reliable and tighter constraints on the holographic dark energy models.
The results of our constraints for the holographic dark energy model without
interaction is $c=0.748^{+0.108}_{-0.009}$, $\Omega_{\mathrm{m0}%
}=0.276^{+0.017}_{-0.016}$, and for model with interaction ($%
c=0.692^{+0.135}_{-0.107}$, $\Omega_{\mathrm{m0}}=0.281^{+0.017}_{-0.017}$ ,$%
\alpha=-0.006 ^{+0.021}_{-0.024}$, where $\alpha$ is an interacting
parameter). As these models have more parameters than the $\Lambda$CDM
model, we use the Bayesian evidence as a model selection criterion to make
comparison. We found that the holographic dark energy models are mildly
favored by the observations compared with the $\mathrm{\Lambda CDM}$ model.
\end{abstract}

\maketitle


\section{Introduction}

It has been realized that our Universe is experiencing an accelerated
expansion, as shown by several astronomical observations \cite{SN}. The
acceleration of the Universe strongly indicates the existence of a
mysterious exotic matter, namely the dark energy, which has large enough
negative pressure and makes up the largest portion of the total matter in
the current Universe. The combined analysis of observational data suggests
that the Universe is spatially flat, and consists of approximately 3/4 dark
energy, 1/4 dust matter (cold dark matter plus baryons), and negligible
amount of radiation \cite{SN,LSS,CMB}. The simplest candidate of dark energy
is the cosmological constant $\Lambda $ (vacuum energy)
which has the equation of state $w=-1$. The cosmological constant-cold dark
matter model ($\Lambda $CDM) works very well, and is in agreement with a
large number of resent observations. However, there are two problems in this
scenario --- the \textit{fine-tuning} problem and the \textit{cosmic
coincidence} problem \cite{coincidence}. The fine-tuning problem asks why
the vacuum energy density today is so small ($\mathrm{10^{-47}GeV^{4}}$)
compared with the theoretical value ($\mathrm{10^{74}GeV^{4}}$) from the
quantum gravity. The cosmic coincidence problem is that since the evolution
of the energy densities of dark matter and dark energy are so different
during the expansion of the Universe, why are they nearly equal to each
other today?

The dark energy may be a problem which has to be solved in the context of
quantum gravity \cite{Witten:2000zk}. In the classical gravity theory, the
dark energy density (cosmological constant) can be an arbitrary value.
However, a complete theory of quantum gravity should be capable of
determining the properties of dark energy such as the energy density and the
equation of state \cite{Witten:2000zk}. The holographic dark energy model is
an attempt to apply the holographic principle of quantum gravity theory to
the dark energy problem \cite%
{Cohen:1998zx,Horava:2000tb,Hsu:2004ri,Li:2004rb}.

It is well known that the holographic principle is an important result from
the explorations of the quantum gravity theory and string theory \cite%
{holoprin}, enlightened by investigations on the properties of black holes.
For an effective field theory in a box of size $L$ with UV cut-off $\Lambda
_{c}$, the entropy $S$ scales extensively as $S\sim L^{3}\Lambda _{c}^{3}$.
However, considering the peculiar thermodynamics of black hole \cite{bh},
the maximum entropy in a box of volume $L^{3}$ may behave nonextensively,
growing only as the surface area of the box, i.e.$S\leq S_{\text{BH}}\equiv
\pi M_{\mathrm{pl}}^{2}L^{2}$ (Bekenstein entropy bound). This nonextensive
scaling suggests that quantum field theory breaks down in large volume. To
reconcile this breakdown with the success of local quantum field theory,
Cohen et al. \cite{Cohen:1998zx} proposed a more restrictive energy bound.
They pointed out that in quantum field theory a short distance (UV) cut-off
is related to a long distance (IR) cut-off due to the limit set by forming a
black hole. In other words, if the quantum zero-point energy density $\rho
_{\Lambda }$ is relevant to a UV cut-off, the total energy of the whole
system with size $L$ should not exceed the mass of a black hole of the same
size, thus we have $L^{3}\rho _{\Lambda }\leq LM_{\mathrm{pl}}^{2}$. When we
take the whole Universe into account, the vacuum energy related to this
holographic principle \cite{holoprin} is viewed as dark energy, usually
dubbed holographic dark energy. The largest IR cut-off $L$ is chosen by
saturating the inequality so that we get the holographic dark energy density
\begin{equation}
\rho _{\mathrm{de}}=3c^{2}M_{\mathrm{pl}}^{2}L^{-2}~,  \label{Holo}
\end{equation}%
where $c$ is a numerical constant, and $M_{\mathrm{pl}}\equiv 1/\sqrt{8\pi G}
$ is the reduced Planck mass. If we take $L$ as the size of the current
Universe, for instance the Hubble scale $H^{-1}$, then the dark energy
density will be close to the observational result. However, Hsu \cite%
{Hsu:2004ri} pointed out that this yields a wrong equation of state for dark
energy. Li \cite{Li:2004rb} subsequently proposed that the IR cut-off $L$
should be taken as the size of the future event horizon
\begin{equation}
R_{\text{\textrm{eh}}}(a)=a\int_{t}^{\infty }{\frac{dt^{\prime }}{%
a(t^{\prime })}}=a\int_{a}^{\infty }{\frac{da^{\prime }}{Ha^{\prime 2}}}~.
\label{Horizon}
\end{equation}%
Then this problem can be solved, and the holographic model can thus be
constructed successfully. For extensive studies of this model (HDE), see
Ref. \cite%
{Huang,Enqvist04,Ke,HuangLi,Zhang,Pavon,WangGong,Kim,Nojiri,Hu,Chen,Ma}. The
holographic dark energy may also have interaction with matter, this is the
so called interacting dark energy model (IHDE) \cite{Pavon,WangGong}. Both
the HDE and IHDE model can be tested by cosmological observations, such as
the type Ia supernovae \cite{HuangGong}, CMB \cite{Enqvist05,Shen,Kao},
X-ray gas mass fraction in the clusters \cite{Chang}, differential ages of
passively evolving galaxies \cite{Yi} and combinations of SN Ia, CMB and LSS
data \cite{ZhangWu,WangLin,WuGong,Feng}. Currently, the tightest constraint
on the HDE model is given by \cite{WuGong}, namely $c=0.85^{+0.18}_{-0.02}$
and $\Omega_{m0}=0.27^{+0.04}_ {-0.03}$, but $c>1$ is still allowed under
this constraint. In addition, by using the $\chi ^{2}$ statistic as a model
comparison technique, Ref.\cite{WuGong} suggests that HDE model is equally
favored by the current observational data compared with the $\Lambda$CDM
model.

In this paper we consider the dynamical behavior of the holographic dark
energy models, particularly what is the condition for the Universe to end in
the so called \textquotedblleft Big Rip\textquotedblright\ \cite{bigrip}. we
then use several new data sets to constraint the holographic dark energy
models. These includes a sample of 182 high quality SN Ia data (Despite
coincidence in number, this is not exactly the same data set used by Gold06
\cite{Riess06}), a sample of 42 latest X-rays gas mass fraction ($\mathrm{%
f_{gas}}$) data \cite{Allen07}, a sample of 27 gamma-ray burst (GRB) data%
\cite{Schaefer07} generated from one of the tightest correlations---the $%
E_{peak}-E_{\gamma }$ correlation \cite{Ghirlanda04}, the baryon acoustic
oscillation measurement from the Sloan Digital Sky Survey\cite{Eisenstein05}%
, and the CMB shift parameter from WMAP 3 years result. With these more
diverse set of observations, we can obtain constraints which are not only
tighter, but also more reliable.

We compare the observational fit of the HDE and IHDE models with the $%
\Lambda $CDM model. In previous works, the $\chi ^{2}$ statistic has been
used for comparison. However, as the holographic dark energy models (with or
without interaction) have more parameters than the $\Lambda$CDM, the $\chi^2$
may not be a suitable criterion for making comparison. Here we use the
Bayesian Evidence (BE) as the information criterion to assess the strength
of the holographic models.

The paper is organized as follows: In \S 2, we review our models of
holographic dark energy, and derive the evolution equations. We also discuss
qualitatively the behavior of the HDE and IHDE models, particularly the
impact of the interacting term on the fate of the Universe. In \S 3, we
present our method of analysis and the data used (\S 3A), then briefly
discuss the Markov Chain Monte Carlo (MCMC) techniques (\S 3B) and the model
selection criteria (\S 3C). In \S 4, we show the results of our constraints,
and using the Bayesian evidence we make comparisons between the holographic
dark energy models (including the HDE and IHDE) and the $\Lambda $CDM model.
We summarize our result and conclude in \S 5.

\section{The models}

For a spatially flat (the flat geometry is assumed throughout this paper)
Friedmann-Robertson-Walker (FRW) Universe filled with matter component $\rho
_{\mathrm{m}}$ and holographic dark energy $\rho _{\mathrm{de}}$, the
Friedmann equation reads
\begin{equation}
3M_{\mathrm{pl}}^{2}H^{2}=\rho _{\mathrm{m}}+\rho _{\mathrm{de}}~.
\label{Friedmann}
\end{equation}%
where $\rho _{\mathrm{de}}=3c^{2}M_{\mathrm{pl}}^{2}L^{-2}$, $L$ is the
future event horizon, i.e. $L(t)=R_{\text{\textrm{eh}}}(a)$.

When we consider the interaction team between the two dark components, the
conservation equations can be written as%
\begin{equation}
\dot{\rho}_{m}+3H\rho _{m}=Q,  \label{Matter}
\end{equation}%
\begin{equation}
\dot{\rho}_{\mathrm{de}}+3H(1+w_{\mathrm{de}})\rho _{\mathrm{de}}=-Q,
\label{Dark Energy}
\end{equation}%
where $w_{\mathrm{de}}$ is the equation of state of holographic dark energy.
The form of the interaction $Q$ is not unique. Here, we consider an
interaction term of the following form (similar to but slightly different
from those given in \cite{Pavon}):
\begin{equation}
Q=3\alpha H\rho _{\mathrm{de}},  \label{Interaction}
\end{equation}%
This form of interaction has been discussed in some literatures on inflation
and reheating. For example, in the warm inflationary model \cite{Ferreira},
in which the scalar field's energy is transferred to the matter due to
scalar field oscillations, there is an interaction term which is throughout
the inflationary regime (not just after slow-roll), so that the energy of
the scalar field is transferred to the matter content continuously and the
matter content is not driven to zero \cite{Billyard}. Also, in string
theory, a similar interaction term arises in the Einstein frame which
depends on the dark energy density. Moreover, a term of the form (\ref%
{Interaction}) could be motivated by analogy with dissipation, for instance,
a fluid with bulk viscosity may give rise to a term of this form in the
conservation equation \cite{Billyard,Eckart}. We note, however, that
interactions of other form is also possible and has been discussed in the
literature \cite{BCLM08,K08,Miao08}

Using the definition of holographic dark energy (\ref{Holo}) and the
relationship $\rho_{de}=3H^{2}M_{pl}^{2}\Omega_{de}$ (in which $\Omega_{de}$%
is the fractional dark energy density), taking the derivative with respect
to $x=\ln a$, we obtain
\begin{equation}
\rho _{\mathrm{de}}^{^{\prime }}\equiv \frac{d\rho _{\mathrm{de}}}{dx}=-6M_{%
\mathrm{pl}}^{2}H^{2}\Omega _{\mathrm{de}}(1-\frac{\sqrt{\Omega _{\mathrm{de}%
}}}{c}).  \label{Derivative}
\end{equation}%
Considering the derivative relationship between $t$ and $z$: $\frac{d}{dt}=H%
\frac{d}{dx}=-H(1+z)\frac{d}{dz}$ and Eq. (\ref{Dark Energy}) and (\ref%
{Interaction}), we have the following equation
\begin{equation}
\rho _{\mathrm{de}}^{^{\prime }}+3(1+w_{\mathrm{de}})\rho _{\mathrm{de}%
}=-3\alpha \rho _{\mathrm{de}}.  \label{Simple Derivative}
\end{equation}

After taking derivative about $\rho _{\mathrm{de}}=3H^{2}M_{\mathrm{pl}%
}^{2}\Omega _{\mathrm{de}}$ and substitute Eq.(\ref{Derivative}) into it, we
obtain%
\begin{equation}
\frac{H^{^{\prime }}}{H}=-\frac{\Omega _{\mathrm{de}}^{^{\prime }}}{2\Omega
_{\mathrm{de}}}+\frac{\sqrt{\Omega _{\mathrm{de}}}}{c}-1.  \label{Hubble 1}
\end{equation}%
On the other hand, using the Friedmann equation $\dot{H}=-\frac{1}{2M_{%
\mathrm{pl}}^{2}}(\rho +p)$($\rho$ and $p$ are the total energy density and
pressure), and substitute $\dot{H}=H^{^{\prime }}H$ and $w_{\mathrm{de}}$
from Eq. (\ref{Simple Derivative}) into it, we could get
\begin{equation}
\frac{H^{^{\prime }}}{H}=\frac{1}{2}\Omega _{\mathrm{de}}-\frac{3}{2}+\frac{1%
}{c}\Omega _{\mathrm{de}}^{\frac{3}{2}}+\frac{3}{2}\alpha \Omega _{\mathrm{de%
}}.  \label{Hubble 2}
\end{equation}%
Combining Eq. (\ref{Hubble 1}) and (\ref{Hubble 2}), we find the
differential equation for $\Omega _{\mathrm{de}}$
\begin{equation}
\frac{d\Omega _{\mathrm{de}}(z)}{dx}=\Omega _{\mathrm{de}}[(1-\Omega _{%
\mathrm{de}})(1+\frac{2}{c}\sqrt{\Omega _{\mathrm{de}}})-3\alpha \Omega _{%
\mathrm{de}}],  \label{Density 1}
\end{equation}%
i.e.
\begin{equation}
\frac{d\Omega _{\mathrm{de}}(z)}{dz}+\frac{\Omega _{\mathrm{de}}}{1+z}%
[(1-\Omega _{\mathrm{de}})(1+\frac{2}{c}\sqrt{\Omega _{\mathrm{de}}}%
)-3\alpha \Omega _{\mathrm{de}}]=0.  \label{Density 2}
\end{equation}%
Consequently, the differential equation for Hubble parameter $H(z)$ could be
written as
\begin{equation}
\frac{dH}{dz}=-\frac{H(z)}{1+z}[\frac{1}{2}\Omega _{\mathrm{de}}(1+3\alpha +%
\frac{2}{c}\sqrt{\Omega _{\mathrm{de}}})-\frac{3}{2}].  \label{Hubble 3}
\end{equation}%
Eq.(\ref{Density 2}) and (\ref{Hubble 3}) can be solved numerically to
obtain the expansion rate $H(z)$.

We now study the behavior of the holographic dark energy models. The
evolution of dark energy can be understood by inspecting its effective
equation of state, for which we have
\begin{equation}
\rho _{de}^{\prime }+3(1+w_{\mathrm{eff}})\rho _{\mathrm{de}}\equiv 0.
\end{equation}%
Using Eqs.(\ref{Derivative}) and (\ref{Simple Derivative}) we obtain
\begin{equation}
w_{\mathrm{eff}}(z)=w_{de}+\alpha =-\frac{1}{3}-\frac{2}{3}\frac{\sqrt{%
\Omega _{\mathrm{de}}}}{c}.  \label{EEoS}
\end{equation}%
It is interesting to note here that $\alpha $ (i.e. the interaction rate of
dark energy and dark matter) does NOT show up explicitly in this equation,
but only implicitly affects the result through its effect on $\Omega _{%
\mathrm{de}}$, so this equation applies to both HDE and IHDE models. For the
flat Universe with dark energy, the dark energy would eventually dominate
the density, $\Omega _{\mathrm{de}}\rightarrow 1$, hence the effective
equation of state of the holographic dark energy evolves dynamically, and
\begin{equation}
w_{\mathrm{eff}}\rightarrow -\frac{1}{3}-\frac{2}{3c}.  \label{eq:wefflim}
\end{equation}

It is obvious from Eqs.~(\ref{EEoS})-(\ref{eq:wefflim}) that for the HDE
model, the equation of state depends on the parameter $c$. For $c>1$, the
equation of state $w>-1$, and the holographic dark energy behaves as
quintessence. For $c<1$, $w<-1$ could be realized. For such models, the
Universe would end in a Big Rip.

\begin{figure}[t]
\centerline{\includegraphics[bb=0 0 336
225,width=3.6in,height=3.0in]{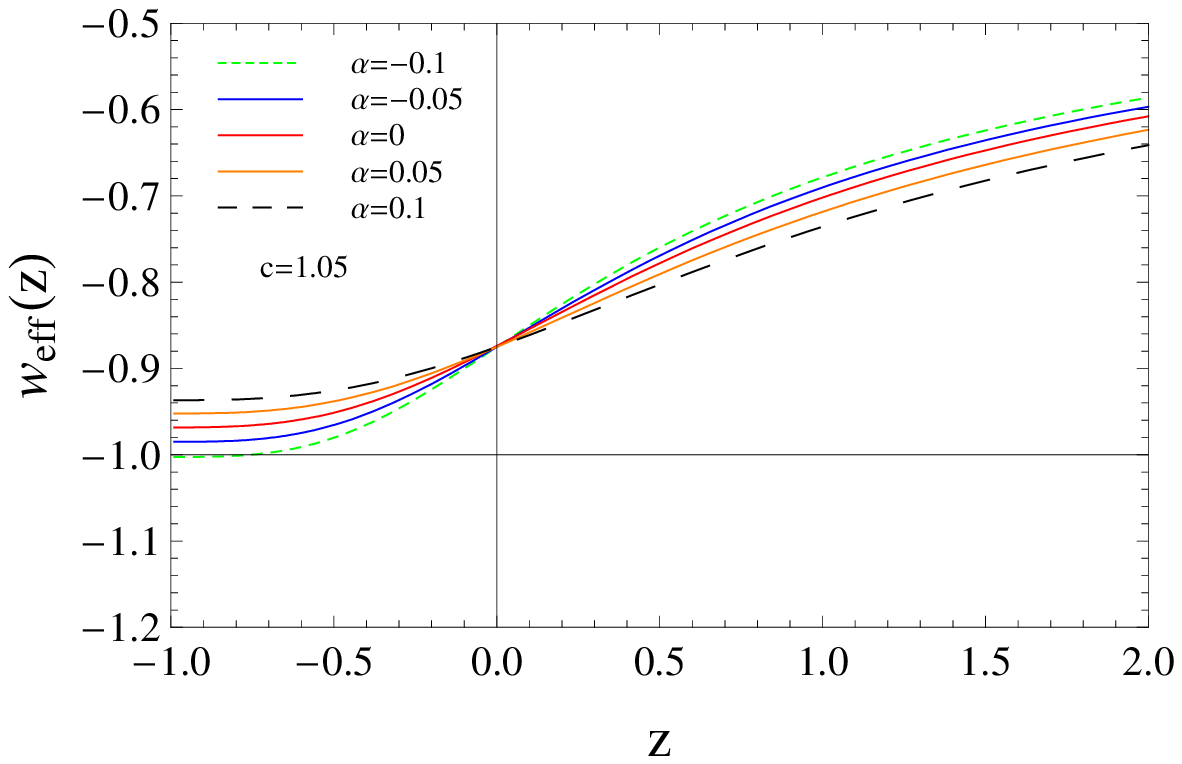}
\includegraphics[bb=6 0 345 227,width=3.6in,height=3.0in]{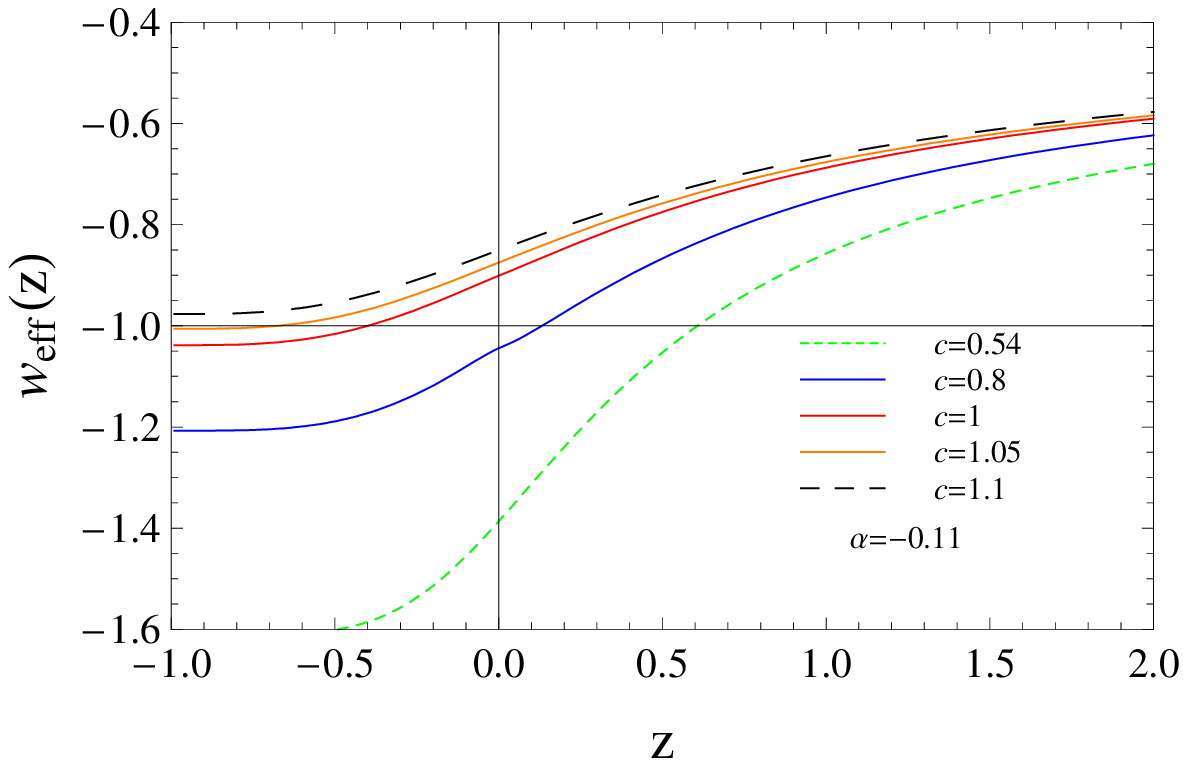}
\hskip 0.3in}
\caption{Equation of state for the selected values of $(c,\protect\alpha )$
or the illustration purpose. Here we set $\Omega _{m0}=0.27$.}
\label{weffihde1}
\end{figure}

For IHDE models, we illustrate the behavior of the model by choosing some
representative values of $\alpha$ and c, and plot the evolution of equation
of state $w(z)$ in Fig. \ref{weffihde1}. On the left of Fig. \ref{weffihde1}%
, we show the evolution of equation of state $w $ as a function of $z$ for
fixed value of $c$ ($c=1.05$) and different values of $\alpha $. Initially,
all models have $-1<w_{eff}<0$. If $\alpha =0$, the model is reduced to the
case of HDE, for which the equation of state $w \to -1$ (cosmological
constant) as $z\to -1$ ($a \to \infty$). For $\alpha >0$, energy is
transferred from dark energy to dark matter, making the effective equation
of state of the dark energy greater than $-1$, so the dark energy behaves as
quintessence-like. For $\alpha <0$, energy is transferred from dark matter
to dark energy, so the effective equation of state of dark energy crosses $%
-1 $, i.e. it exhibits quintom behaviors.
On the right side of Fig.~\ref{weffihde1}, we plot the equation of state for
fixed $\alpha $ with different values of $c$ ($\alpha =-0.11$ in this
particular case). As $c$ increases, the equation of state depart from $-1$
and increases. For large value of $c$ the quintom divide is not crossed, and
the model again behaves like a quintessence.
\begin{figure}[t]
\centerline{\includegraphics[bb=6 0 350
230,width=3.4in,height=2.9in]{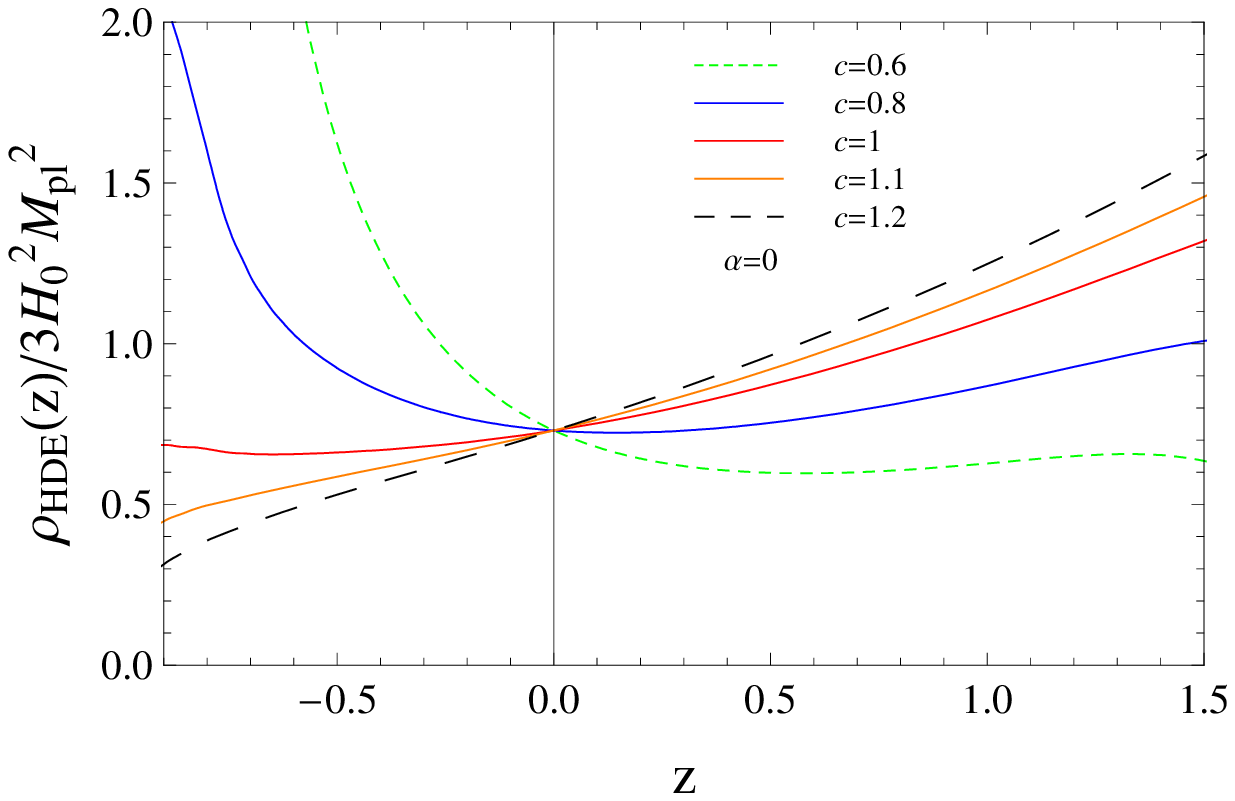}
\includegraphics[bb=6 0 350 232,width=3.4in,height=2.9in]{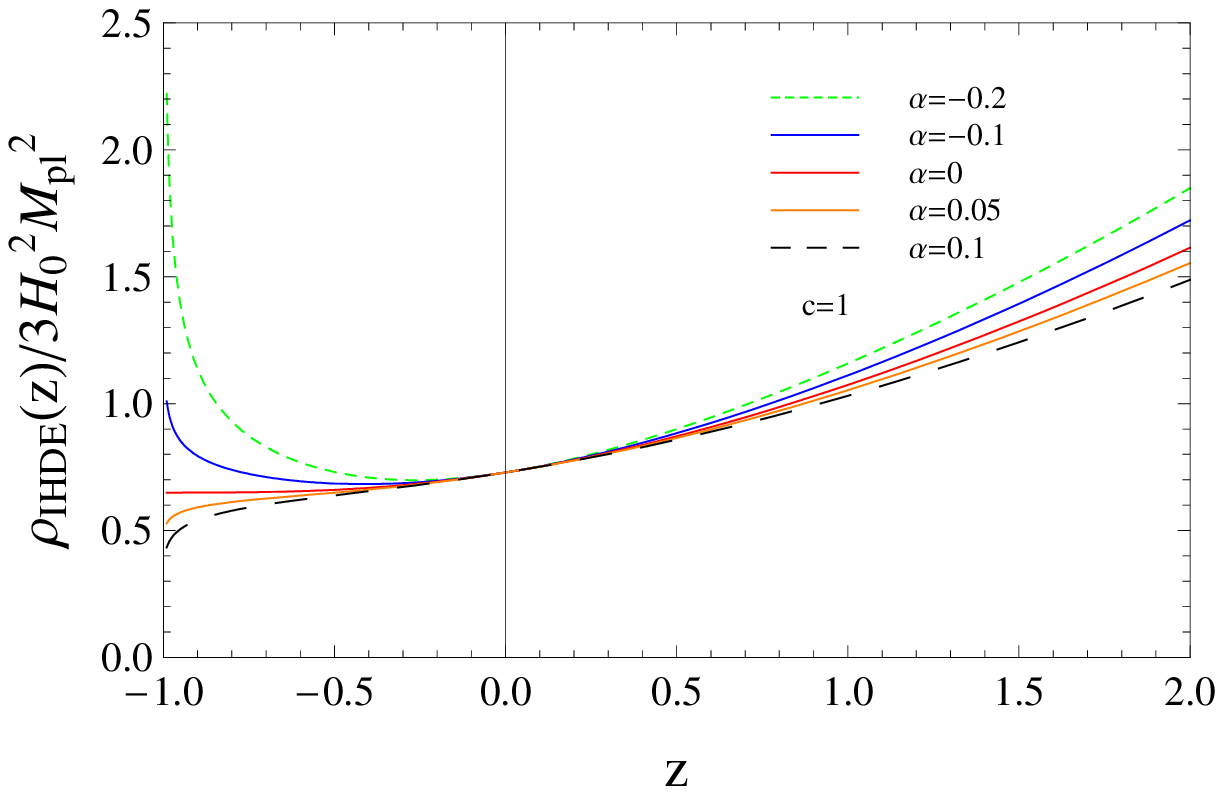}
\hskip 0.3in}
\caption{Energy density of IHDE and HDE model. Left: Energy density of HDE .
Right: Energy density of IHDE. Here we set $\Omega _{m0}=0.27$.}
\label{rho}
\end{figure}
In Fig.~\ref{rho} we plot $\rho _{\mathrm{de}}(z)$. On the left of Fig.~\ref%
{rho}, we plot the special case of HDE ($\alpha=0$). we find that for $0<c<1
$, the equation of state will cross $-1$ and behave like a ``quintom'', so
that $\rho_{\mathrm{HDE}}$ diverges in finite time; for $c>1$, the equation
of state is greater than $-1$, and the dark energy density decrease as time
passes. On the right of Fig. \ref{rho}, we plot behavior of the system with
different values of $\alpha$ with $c=1$. For sufficiently negative $\alpha$,
the IHDE density diverges as energy is transferred from the dark matter to
dark energy, and eventually we would run into the ``Big Rip''.

\section{METHODOLOGY}

\subsection{Data analysis}

We utilize several data sets to constrain the parameters of the holographic
dark energy model, including a selection of 182 high-quality type Ia
supernovae, the baryon acoustic oscillation measurement from the Sloan
Digital Sky Survey, 42 latest X-ray gas mass fraction data from Chandra
observations, 27 GRB samples generated with $E_{peak}-E_{\gamma }$
correlation, and the CMB shift parameter from WMAP 3 years result.

\subsubsection{Selected SN Ia data set}

For the SN Ia data, our sample includes 77 highest quality data from the
Riess Gold 06 sample \cite{Riess06} (30 HST supernovae and 47 high quality
SNLS supernovae \cite{Astier05})\cite{Nesseris06}, and 105 ESSENCE
supernovae \cite{Wood07} (60 observed by the ESSENCE team and 45 nearby
supernovae re-analyzed by the ESSENCE team), the total number is 182. We use
the light curve fitters MLCS2k2 \cite{Riess96,Jha06} (the other light curve
fitter SALT \cite{Guy05} give results which are consistent with MLCS2k2 \cite%
{Riess06, Wood07}), this algorithm avoids the need of normalization \cite%
{Davis07}. The redshift of this data set reaches 1.755.

The likelihood function of the parameters can be determined from $\chi ^{2}$
statistics, and for SN Ia data
\begin{equation}
\chi _{\mathrm{SN_{sel}}}^{2}(\mathbf{\theta })=\sum_{i=1}^{182}\frac{(\mu
_{obs}(z_{i})-\mu _{th}(z_{i}))^{2}}{\sigma _{i}^{2}},  \label{eq:chisq}
\end{equation}%
where the theoretical value of distance modulus $\mu _{th}(z)$ is given by
\begin{eqnarray}
\mu _{th}(z) &=&5\log _{10}d_{L}(z)+25  \notag \\
&=&5\log _{10}D_{L}(z)-5\log _{10}h_{0}+42.38,  \label{mut}
\end{eqnarray}%
and
\begin{equation}
D_{L}(z)=\frac{H_{0}}{c}\times d_{L}(z).  \label{LD}
\end{equation}%
Here $H_{0}=100h_{0}\mathrm{kms}^{-1}\mathrm{Mpc}^{-1}$ and the luminosity
distance $d_{L}$ can be written as
\begin{equation}
d_{L}(z)=(1+z)\int_{0}^{z}\frac{cdz^{\prime }}{H(z^{\prime })}.
\label{fo:dl}
\end{equation}

\subsubsection{BAO measurement from SDSS}

The baryon acoustic oscillation signatures in the large-scale clustering of
galaxies can be seen as a standard ruler providing another way to explore
the expansion history of the Universe. We use the measurement of the BAO
peak from a spectroscopic sample of 46,748 luminous red galaxies (LRGs)
observations of SDSS to test cosmology \cite{Eisenstein05}, which gives the
value of $\mathrm{A}=0.469(n_{s}/0.98)^{-0.35}\pm 0.017$ at $z_{\mathrm{BAO}%
}=0.35$ where $n_{s}=0.95$ \cite{Spergel06}. The expression of A can be
written as
\begin{equation}
A=\frac{\sqrt{\Omega _{m0}}}{(H(z_{\mathrm{BAO}})/H_{0})^{1/3}}\left[ \frac{1%
}{z_{\mathrm{BAO}}}\int_{0}^{z_{\mathrm{BAO}}}\frac{dz^{\prime }}{%
H(z^{\prime })/H_{0}}\right] ^{2/3}\mathrm{,}  \label{A}
\end{equation}%
\newline
and the $\chi _{\mathrm{BAO}}^{2}$ is
\begin{equation}
\chi _{\mathrm{BAO}}^{2}=\Bigg(\frac{\mathrm{A}-0.469(n_{s}/0.98)^{-0.35}}{%
0.017}\Bigg)^{2}.  \label{BAO}
\end{equation}

\subsubsection{Latest X-ray gas mass fraction data from Chandra}

The X-ray gas mass fraction in the largest, X-ray luminous, dynamically
relaxed clusters of galaxies provides a fair sample of the matter content of
the Universe. It could give a constraint to the geometry of the Universe
with the relation $\mathrm{f_{gas}\propto d_{A}^{1.5}}$, under the
assumption that this fraction should be approximately constant with redshift
\cite{Allen02, Allen04, Feng}. Here we use the latest $\mathrm{f_{gas}}$
data derived from 42 relaxed clusters by Allen \emph{et. al }\cite{Allen07}.
from $Chandra$ observations \cite{Allen07}, the redshift of this sample
ranges from 0.05 to 1.1.

Following Allen \emph{et. al} \cite{Allen07}, the $\chi _{f_{\mathrm{gas}%
}}^{2}$ is given by
\begin{eqnarray}
\chi _{f_{\mathrm{gas}}}^{2}(\mathbf{\theta }) &=&\Bigg(\sum_{i=1}^{42}\frac{%
[f_{\mathrm{gas}}^{\mathrm{\Lambda CDM}}(z_{i})-f_{\mathrm{gas,i}}]^{2}}{%
\sigma _{f_{\mathrm{gas,i}}}^{2}}\Bigg)+\Bigg(\frac{\Omega
_{b}h_{0}^{2}-0.0214}{0.0020}\Bigg)^{2}+\Bigg(\frac{h_{0}-0.72}{0.08}\Bigg)%
^{2}  \notag \\
&&+\Bigg(\frac{s_{0}-0.16}{0.048}\Bigg)^{2}+\Bigg(\frac{K-1.0}{0.1}\Bigg)%
^{2}+\Bigg(\frac{\eta -0.214}{0.022}\Bigg)^{2},  \label{fgas}
\end{eqnarray}%
and the model fitted to the reference $\Lambda $CDM $\mathrm{(}\Omega
_{m0}=0.3,\Omega _{\Lambda 0}=0.7\mathrm{)}$ data is
\begin{equation}
f_{\mathrm{gas}}^{\Lambda \text{CDM}}(z)=\frac{KA\gamma b(z)}{1+s(z)}\left(
\frac{\Omega _{\mathrm{b}}}{\Omega _{\mathrm{m}}}\right) \left[ \frac{d_{%
\mathrm{A}}^{\Lambda \text{CDM}}(z)}{d_{\mathrm{A}}(z)}\right] ^{1.5},
\label{eq:fgasl}
\end{equation}%
where $d_{A}(z)$ and $d_{A}^{\Lambda \text{CDM}}(z)$ are the angular
diameter distances to the clusters in the test model and reference model,
\begin{equation}
d_{A}(z)=\frac{1}{(1+z)}\int_{0}^{z}\frac{cdz^{\prime }}{H(z^{\prime })}.
\label{AD}
\end{equation}

The parameter $b(z)=b_{0}(1+\alpha _{b}z)$ in Eq.~(\ref{eq:fgasl}) is the
bias factor which parameterizes the redshift-dependent deviation of the
baryon fraction measured at $r_{2500}$ from the Universe mean with $%
0.65<b_{0}<1.0$, $-0.1<\alpha _{b}<0.1$; The factor $s(z)=s_{0}(1+\alpha
_{s}z)$ models the baryonic mass fraction in stars, and $s_{0}=(0.16\pm
0.05)h_{70}^{0.5}$, $-0.2<\alpha _{s}<0.2$. The $A$ accounts for the change
in angle subtended by $r_{2500}$ as the reference cosmological model is
varied:
\begin{equation}
A=\left( \frac{\theta _{2500}^{\Lambda \text{CDM}}}{\theta _{2500}}\right)
^{\eta }\approx \left( \frac{H(z)d_{\mathrm{A}}(z)~~~~~~~}{\left[ H(z)d_{%
\mathrm{A}}(z)\right] ^{\Lambda \text{CDM}}}\right) ^{\eta },
\label{ParameterA}
\end{equation}%
and $\eta $ is the slope of the $f_{\mathrm{gas}}$ in the region of $%
r_{2500} $ measured in the reference $\Lambda $CDM model, which takes the
value $\eta =0.214\pm 0.022$. The parameter $\gamma $ represents the effect
of non-thermal pressure support in the clusters, which ranges from 1.0 to
1.1; The factor $K$ is a `calibration' constant which accounts for residual
uncertainty in the accuracy of the instrument calibration and X-ray
modeling, and we take $K=1\pm 0.1$.

\subsubsection{Gamma-ray bursts data}

The GRB data may be a good complement to the other observational data \cite%
{Ghirlanda04,Schaefer07,Liang05,Ghirlanda06,Dai04,Xu05,Firmani05,Hooper07},
such as SN Ia data. They have very large redshift distribution and can be
observed at much higher redshift, thus provide an effective way to detected
the evolution of the dark energy.

Our GRB data set is constituted of 27 GRB samples in Ref. \cite{Schaefer07}.
They are generated from the $E_{peak}-E_{\gamma }$ correlation (Ghirlanda
relation \cite{Ghirlanda04}), which is one of the tightest correlations for
GRB. The redshift of this data set reaches 6.29.

The $\chi _{\mathrm{GRB}}^{2}$ takes the form of:
\begin{equation}
\chi _{\mathrm{GRB}}^{2}(\mathbf{\theta })=\sum_{i=1}^{27}\frac{(\mu
_{obs}(z_{i})-\mu _{th}(z_{i}))^{2}}{\sigma _{i}^{2}},  \label{eq:gchisq}
\end{equation}%
\newline
in which a distance modulus $\mu _{obs}(z)$ estimated from the observational
data can be calculated as
\begin{equation}
\mu _{obs}(z)=5~\log _{10}(d_{L_{obs}})+25,  \label{GRB 1}
\end{equation}%
with a estimated luminosity distance $d_{L_{obs}}$ expressed in unit of
megaparsecs, and
\begin{equation}
d_{L_{obs}}=[E_{\gamma }(1+z)/(4\pi F_{beam}S_{bolo})]^{1/2}.  \label{GRB 2}
\end{equation}%
Here $F_{beam}$ is the beaming factor, $S_{bolo}$ is the bolometric fluence
of the burst, and the collimation corrected energy $E_{\gamma }$ can be
fitted by
\begin{equation}
\log E_{\gamma }=a+b\log [E_{peak}(1+z)/300keV],  \label{GRB 3}
\end{equation}%
where $a=50.57$ and $b=1.63$. Its uncertainty is
\begin{equation}
\sigma _{\log E_{\gamma }}^{2}=\sigma _{a}^{2}+(\sigma _{b}\log
[E_{peak}(1+z)/300keV])^{2}+(0.4343b\sigma _{E_{peak}}/E_{peak})^{2}+\sigma
_{E_{\gamma },sys}^{2},  \label{GRB 4}
\end{equation}%
where the $1\sigma $ uncertainties in the intercept and slope are $\sigma
_{a}=0.09$ and $\sigma _{b}=0.03$, and the best estimated $\sigma
_{E_{\gamma },sys}$ is 0.16. The $\sigma _{i}$ in Eq.~(\ref{eq:gchisq}) can
be estimated as
\begin{equation}
\sigma _{i}=[(2.5\sigma _{\log E_{\gamma }})^{2}+(1.086\sigma
_{S_{bolo}}/S_{bolo})^{2}+(1.086\sigma _{F_{beam}}/F_{beam})^{2}]^{1/2}.
\label{GRB 5}
\end{equation}%
At last, the expression of $\mu _{th}(z)$ in Eq.~(\ref{eq:gchisq}) is given
by Eq.~(\ref{mut}).

\subsubsection{CMB shift parameter from WMAP 3 years result}

The shift parameter is a very good complement to the previous data set
because the very large redshift distribution ($z_{CMB}=1089$) can reflect
the evolution of the dark energy. The shift parameter $R$ is derived from
the CMB data takes the form as
\begin{eqnarray}
\mathrm{R = \sqrt{\Omega_{m0}}\int^{z_{CMB}}_0\frac{dz^{\prime }}{%
H(z^{\prime })/H_0}.}
\end{eqnarray}
The WMAP3 data gives $\mathrm{R = 1.70\pm0.03}$ \cite{wang07}, thus we have
\begin{eqnarray}
\chi^2_{\mathrm{CMB}} = \Big(\frac{\mathrm{R-1.70}}{0.03}\Big)^2.
\end{eqnarray}

To break the degeneracy and explore the power and differences of the
constraints for these data sets, we use them in several combinations to
perform our fitting: $\mathrm{SN_{sel}+BAO,SN_{sel}+BAO+f_{gas}}$, and $%
\mathrm{SN_{sel}+BAO+f_{gas}+GRB+CMB}$.

\subsection{MCMC}

Markov Chain Monte Carlo (MCMC) techniques are widely used to generate
random samples to simulate the posterior probability of the parameters given
the data sets. This method has several advantages over grid-based approach.
Most importantly, the computational time cost increases approximately
linearly with the number of parameters, so even for a large number of
parameters the estimate can be done within an acceptable computation time
\cite{Lewis02,book1,book2,Laurence06}.

We use the Metropolis-Hastings algorithm with uniform prior to decide
whether to accept a new point into the chain by an acceptance probability:
\begin{equation*}
\mathbf{a}(\mathbf{\theta _{n+1}|\theta _{n}})=\min \mathbf{\Bigg\{\frac{%
p(\theta _{n+1}|d)\;q(\theta _{n}|\theta _{n+1})}{p(\theta _{n}|d)\;q(\theta
_{n+1}|\theta _{n})}\ ,1\Bigg\},}
\end{equation*}%
where $\mathbf{p(\theta )}$ is the prior probability distribution and $%
\mathbf{q(\theta _{n+1}|\theta _{n})}$ is the proposal density of proposing
a new point $\theta _{n+1}$ given a current point $\theta _{n}$ in the
chain. If $\mathbf{a}=1$, the new point $\theta _{n+1}$ is accepted;
otherwise, the new point is accepted with probability $\mathbf{a}$. The
trials are repeated until a new point is accepted, and then we set $\theta
_{n}=\theta _{n+1}$. In our computation, we set a Gaussian-distributed
proposal density for every point which is independent of the position on the
chain, so that $\mathbf{q(\theta _{n+1}|\theta _{n})}$ and $\mathbf{q(\theta
_{n}|\theta _{n+1})}$ are canceled, and consider the uniform prior and
Bayes' theorem we get
\begin{equation}
\mathbf{a}(\mathbf{\theta _{n+1}|\theta _{n}})=\min \Bigg\{\frac{{\mathcal{L}%
}\mathbf{(d|\theta _{n+1})}}{{\mathcal{L}}\mathbf{(d|\theta _{n})}}\ ,1%
\Bigg\}.  \label{Probability}
\end{equation}%
Here ${\mathcal{L}}(\mathbf{d|\theta })$ is the likelihood to obtain the
data set $\mathbf{d}$ given the parameter set $\mathbf{\theta }$, and
usually can be written as
\begin{equation}
{\mathcal{L}}(\mathbf{d|\theta })=\frac{1}{\sqrt{2\pi }\mathbf{\sigma _{d}}}%
e^{-\frac{1}{2}\mathbf{\chi }^{2}}.  \label{eq:likelihood}
\end{equation}

For our three data sets, the $\chi^2$ are
\begin{eqnarray}
\chi^2 = \left\{%
\begin{array}{ll}
\chi^2_{\mathrm{SN_{sel}}}+\chi^2_{\mathrm{BAO}} &  \\
\chi^2_{\mathrm{SN_{sel}}}+\chi^2_{\mathrm{BAO}}+\chi^2_{\mathrm{f_{gas}}} &
\\
\chi^2_{\mathrm{SN_{sel}}}+\chi^2_{\mathrm{BAO}}+\chi^2_{\mathrm{f_{gas}}%
}+\chi^2_{\mathrm{GRB}}+\chi^2_{\mathrm{CMB}}, &
\end{array}%
\right.
\end{eqnarray}
respectively. We assume uniform prior for the parameters of our models
within the given ranges as following: $\Omega_{m0} \in (0,1)$, $c \in (0, 2)$%
, $\alpha \in (-0.2,0.2)$ and $h_0 \in (0.4,0.9)$. In particular, when we
use $\mathrm{f_{gas}}$ data, the parameters coming from this data set $%
\mathbf{\theta _{data}}=\{~\Omega _{b},~s_{0},~\alpha _{s},~b_{0},~\alpha
_{b},~\eta ,~\gamma ,~K~\}$, are also included in our MCMC fitting process.

The thermodynamical bound $c>\sqrt{\Omega_D}$ is NOT assumed \textit{a prior}%
, as we wish to assess the value of the data fairly.

We generate six chains for each case we study, and about one hundred
thousand points are sampled in each chain. The form of proposal density we
use is described in Ref.~\cite{Gong07}. After the convergence determined by
Gelman and Rubin \cite{Gelman} criterion and thinning the chains, we merge
them into one chain which consists of about 10,000 points used to simulate
the probability distribution of the parameters.

\subsection{Model Comparison}

For comparing different models, one must choose a statistical variable, the $%
\chi _{min}^{2}$ is the simplest one and is widely used. However, for models
with different number of parameters, the comparison using $\chi^2$ may not
be fair, as one would expect that models with more parameters tends to have
lower $\chi^2$.

The Akaike information criterion (AIC) \cite{Akaike74}
\begin{equation}
\mathrm{AIC}=-2\,\ln \mathcal{L}_{max}+2\,k  \label{AIC}
\end{equation}%
includes the penalization of the number of parameters, where $\mathcal{L}%
_{max}$ is the maximum likelihood and $k$ is the number of parameters \cite%
{Liddle06, Biesiada07}. However, the size of the data is not embodied in the
AIC, if there is a large number of data, then the reduction in $\chi ^{2}$
due to the additional parameters may be very large and the $2k$ term in the
AIC smaller could not compensate it \cite{Liddle04}.

The Bayesian information criterion(BIC) \cite{Schwarz78} can be written as
\begin{equation}
\mathrm{BIC}=-2\,\ln \mathcal{L}_{max}+k\,\ln N,  \label{BIC}
\end{equation}%
where N is the number of data. The BIC tends to penalize the number of
parameters too much if given large number of data \cite{Liddle04,
Biesiada07,Kurek07}.

We will use the Bayesian evidence (BE) as a model selection criterion. The
Bayesian evidence of a model $M$ takes the form
\begin{equation}
\mathrm{BE}=\int {\mathcal{L}(\mathbf{d|\theta },M)\mathbf{p}(\mathbf{\theta
}|M)d\mathbf{\theta }},  \label{BE}
\end{equation}%
where $\mathcal{L}(\mathbf{d|\theta },M)$ is the likelihood function given
the model $M$ and parameters $\mathbf{\theta }$, and $\mathbf{p}(\mathbf{%
\theta }|M)$ is the priors of parameters. The BE may be the best model
selection criterion, as it is the average of likelihood of a model over its
prior of the parameter space and automatically includes the penalties of the
number of parameters and data, so it is more direct, reasonable and
unambiguous than the $\chi _{min}^{2}$, AIC and BIC in model selection \cite%
{Liddle06,Liddle061,John01, Trotta05,Mukherjee06,Mukherjee05,Kunz06,Trotta07}
(For a connection between BE and $\chi _{min}^{2}$ analysis, see Ref.~\cite%
{Marshall04}). The logarithm of BE can be used as a guide for model
comparison (Jeffreys 1961), and we choose the $\Lambda $CDM as the reference
model: $\Delta \ln \mathrm{BE}=\ln \mathrm{BE}_{model}-\ln \mathrm{BE}%
_{\Lambda \text{CDM}}$. The strength of the evidence for the model is
considered according to the numerical value of BE:
\begin{equation}
\left\{
\begin{array}{ll}
\Delta \ln \mathrm{BE}<1 & \quad \text{Weak} \\
1<\Delta \ln \mathrm{BE}<2.5 & \quad \text{Significant} \\
2.5<\Delta \ln \mathrm{BE}<5 & \quad \text{Strong to very strong} \\
\Delta \ln \mathrm{BE}>5 & \quad \text{Decisive}%
\end{array}%
\right.
\end{equation}%
We use the nested sampling technique to compute BE \cite%
{Mukherjee06,Skilling}.

\section{RESULTS}

\subsection{The HDE Model}

\begin{figure}[htp]
\centerline{\includegraphics[width=3in]{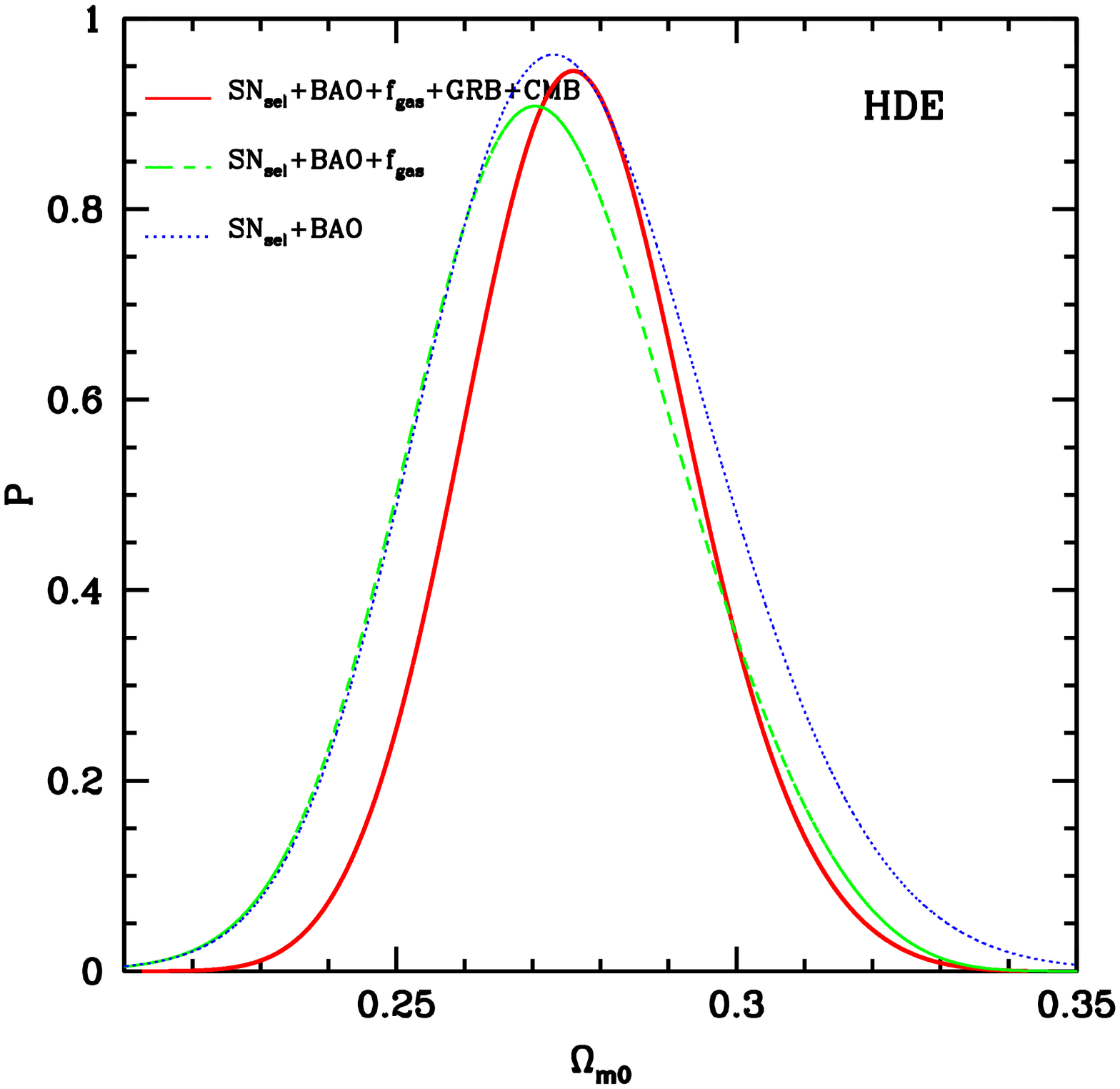}
\includegraphics[width=3in]{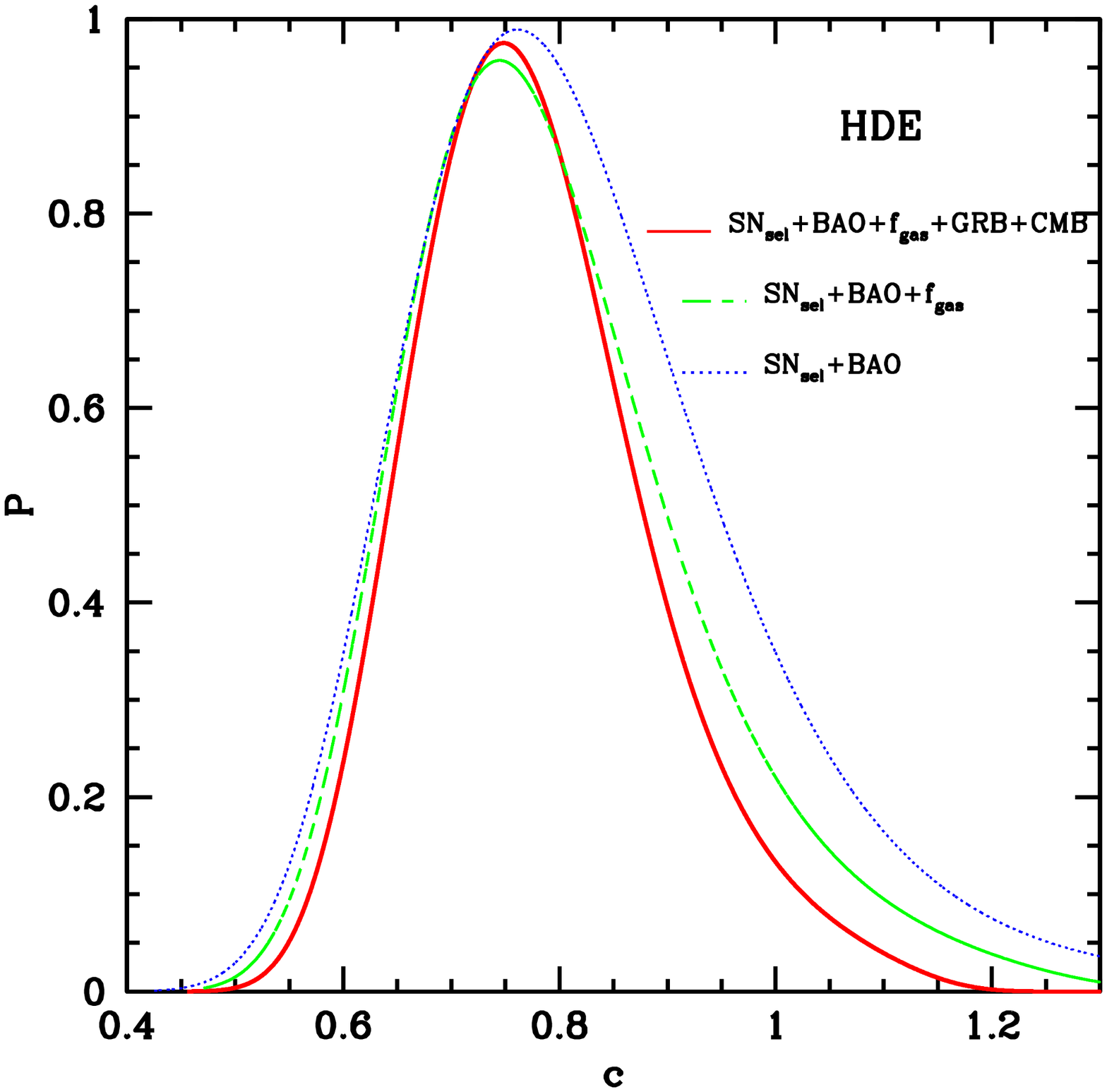}
\hskip 0.3in}
\caption{Probability distribution function (PDF) of the parameters for the
fits of HDE model. Left: The PDF for parameter $\Omega _{m0}$. Right: The
PDF for parameter $c$. }
\label{fig:pdf}
\end{figure}

\begin{table}[hb]
\caption{The fitting result for the HDE model}%
\begin{ruledtabular}
\begin{tabular}[t]{c|c|c|c}
 & $\rm SN+BAO$ & $\rm SN+BAO+f_{gas}$ & $\rm SN+BAO+f_{gas}+GRB+CMB$\\
\hline
$\Omega_{m0}$ & $0.273^{+0.020}_{-0.020}$ & $0.270^{+0.021}_{-0.018}$ & $0.276^{+0.017}_{-0.016}$ \\
$c$ & $0.761^{+0.154}_{-0.117}$ & $0.745^{+0.130}_{-0.101}$ &
$0.748^{+0.108}_{-0.093}$ \\
$\Delta\ln{\rm BE}$ &$~0.09\pm0.12~$ & $~0.63\pm 0.18~$&$~0.65\pm0.18~$\\
\end{tabular}
\end{ruledtabular}
\end{table}
In Table I, we give the best fit value and the $1\sigma $ error of the HDE
model parameters, as well as the value of $\ln \Delta \mathrm{BE}$ for the
three data set combinations. We plot the probability distribution function
(PDF) of the HDE model in Fig.~\ref{fig:pdf}. We can see that the best fit
of $\Omega _{m0}$ are almost the same (around 0.27) for all data sets. The
best fit of $c$ varies slightly across the different data sets, it is 0.761
for the SN+BAO data set, but decreases slightly when the $f_{\mathrm{gas}}$,
GRB and CMB data are included. However, for all data sets, we have $c<1$ at
more than $1.5\sigma $ (see Fig.~\ref{fig:pdf}). The PDF of the parameter
distribution is smoothly distributed.
\begin{figure}[htbp]
\includegraphics[scale = 0.4]{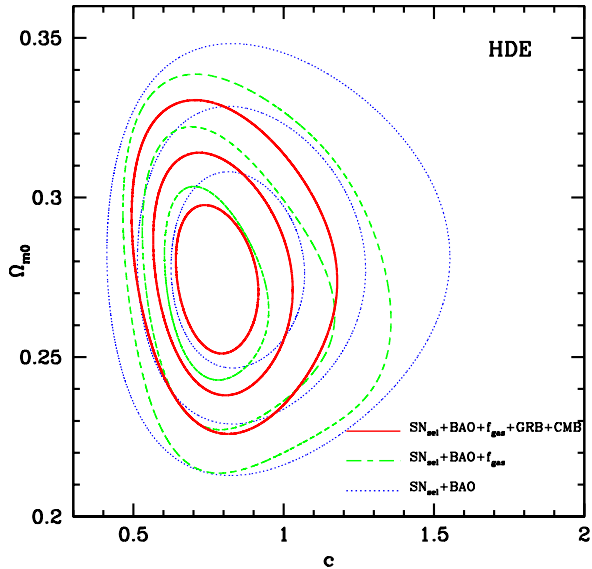}
\caption{The contour maps of c vs. $\Omega _{m0}$ for HDE with $1\protect%
\sigma (68.3\%)$, $2\protect\sigma (95.5\%)$ and $3\protect\sigma (99.7\%)$
confidence levels.}
\label{fig:m0_c}
\end{figure}
The parameter $c$ plays an essential role in determining the evolution of
the HDE model. If $c=1$, the dark energy equation of states would asymptote
to that of a cosmological constant and the Universe would enter the de
Sitter phase in the future; if $c>1$, the equation of state of dark energy
would always be greater than $-1$, it would behave as quintessence dark
energy; if $c<1$, initially the equation of state of HDE would be greater
than -1, but it would decrease and eventually cross the \textquotedblleft
phantom divide line\textquotedblright\ ($w=-1$) as the Universe expands,
acting as a quintom \cite{BFeng}.

\begin{figure}[ht]
\centerline{
\includegraphics[bb=0 0 369 238,width=3.3in,height=2.4in]{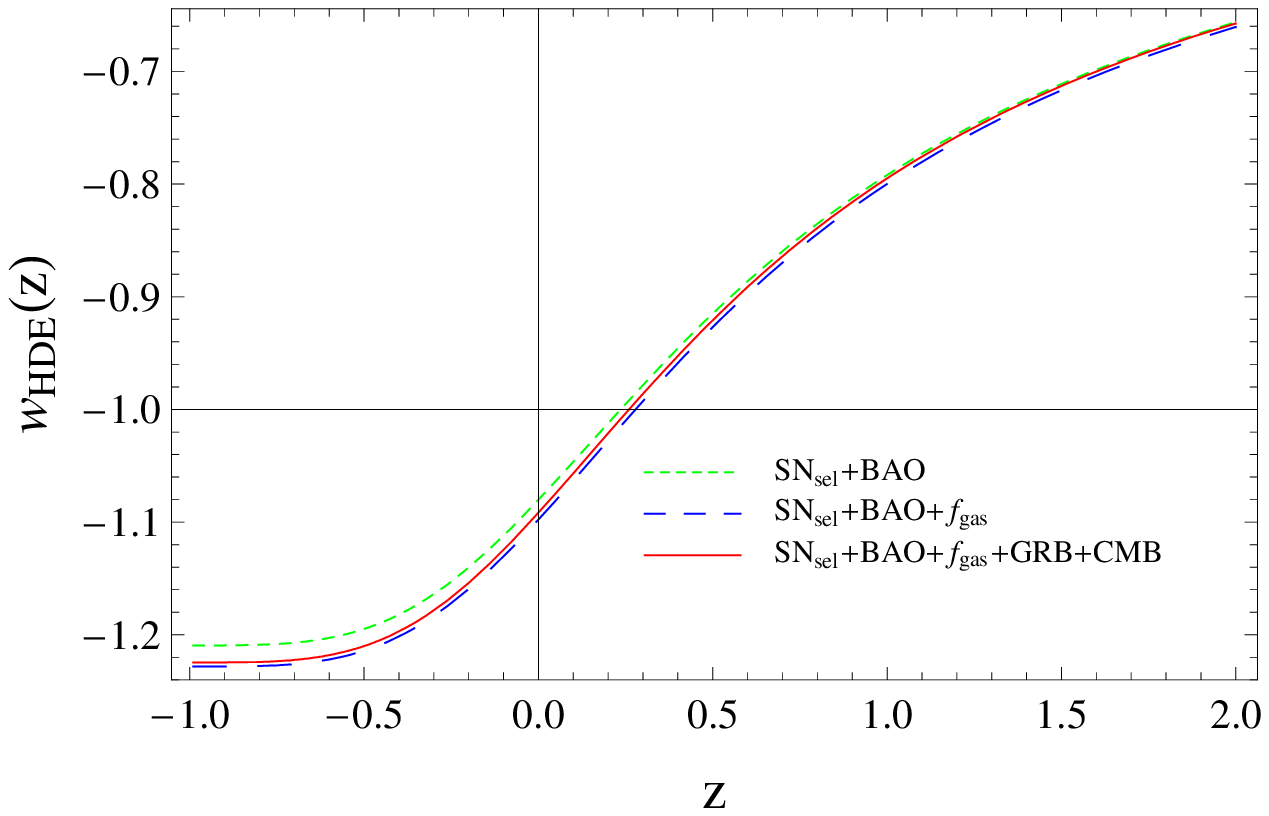}
\includegraphics[bb=0 0 374 245,width=3.2in,height=2.4in]{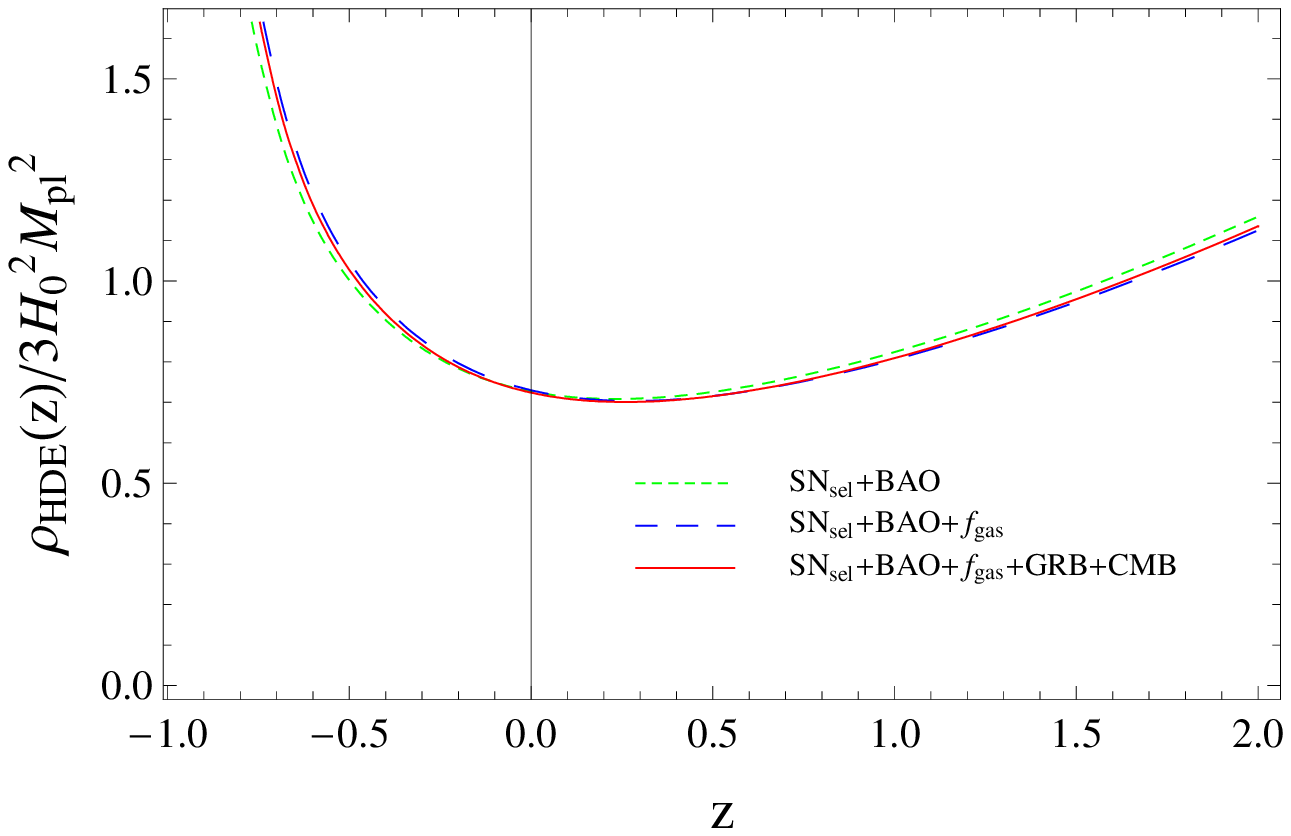}}
\caption{Evolution of the dark energy in the best fit HDE model. Left:
Equation of state. Right: Dark energy density.}
\label{fig:HDE}
\end{figure}

We plot the contour maps of $c$ vs. $\Omega _{m0}$ for HDE with $1\sigma
(68.3\%)$, $2\sigma (95.5\%)$ and $3\sigma (99.7\%)$ confidence levels in
Fig.~\ref{fig:m0_c}. Our constraint is tighter than previous ones, e.g. Ref.~%
\cite{ZhangWu}, as we have used more precise data in our fitting. The center
of the best fit is located at $c<1$, but there is still a fair fraction of
allowed parameter space in which $c>1$. The evolution of $w$ and $\rho _{%
\mathrm{de}}$ in HDE models with the best fit parameters for the three data
set combinations are shown in Fig.~\ref{fig:HDE}. For these cases, as we
expected, the dark energy diverges in finite time and the Universe ends with
a Big Rip.

Moreover, for all the four data sets, we have $c<1.2$ at more than $3\sigma$
(see Fig.~\ref{fig:pdf} and Fig.~\ref{fig:m0_c}), which is rather consistent
with the possible theoretical limit of parameter $c$ from the weak gravity
conjecture (see \cite{Ma}).

The HDE model fits about equally well ($\ln$ BE=0.09) as the $\Lambda$CDM
when we only use the SNIa and BAO data. With $f_{gas}$, GRB and CMB data
added, it fits mildly better than the $\Lambda$CDM, but with the data
presently available the difference is not significant ($\ln$ BE =$0.63\sim
0.65$).

\subsection{The IHDE Model}

\begin{figure}[b]
\centerline{\includegraphics[bb=0 0 410
263,width=3.1in,height=2.1in]{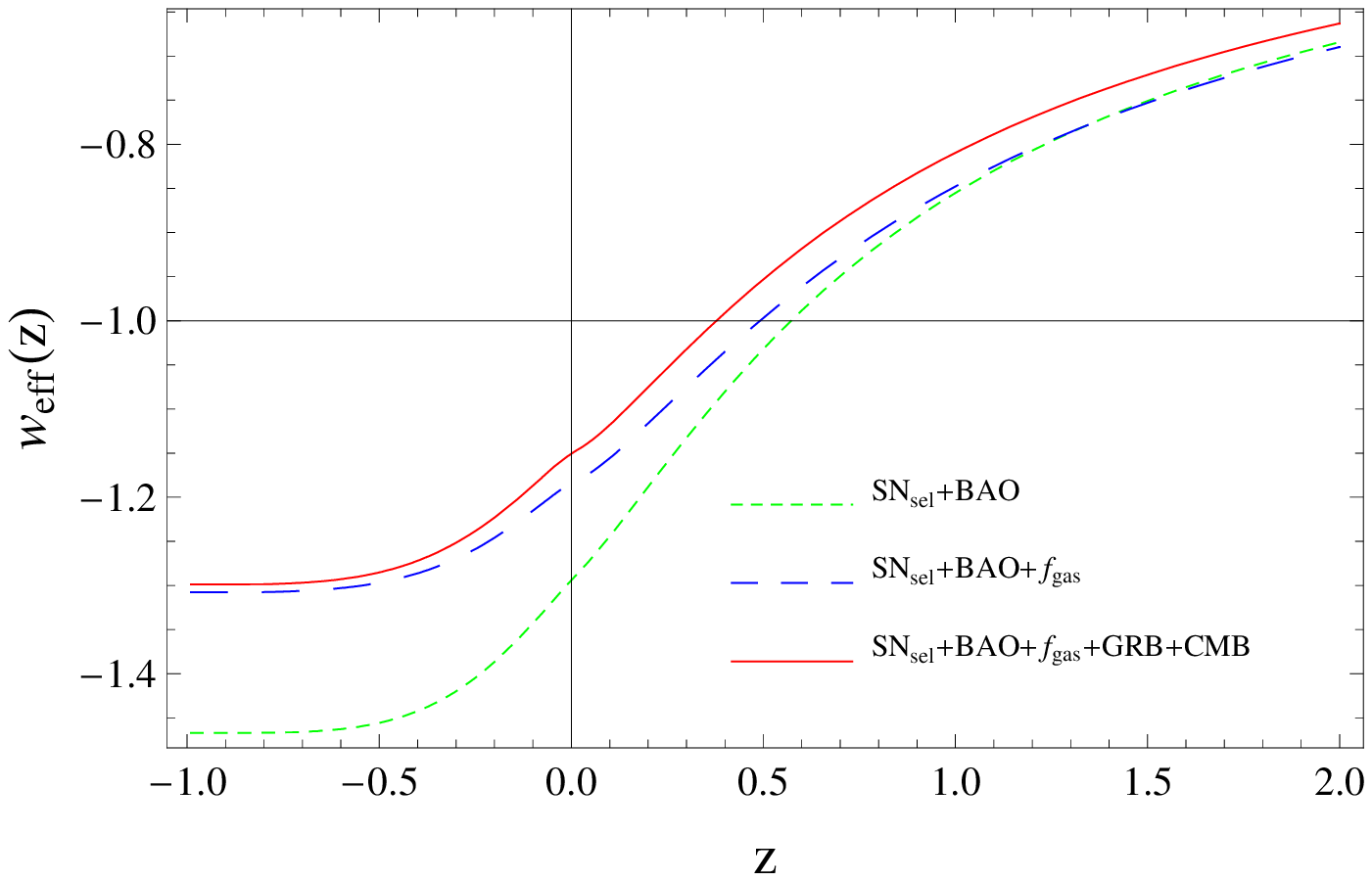}
\includegraphics[bb=0 0 370 241,width=3.1in,height=2.1in]{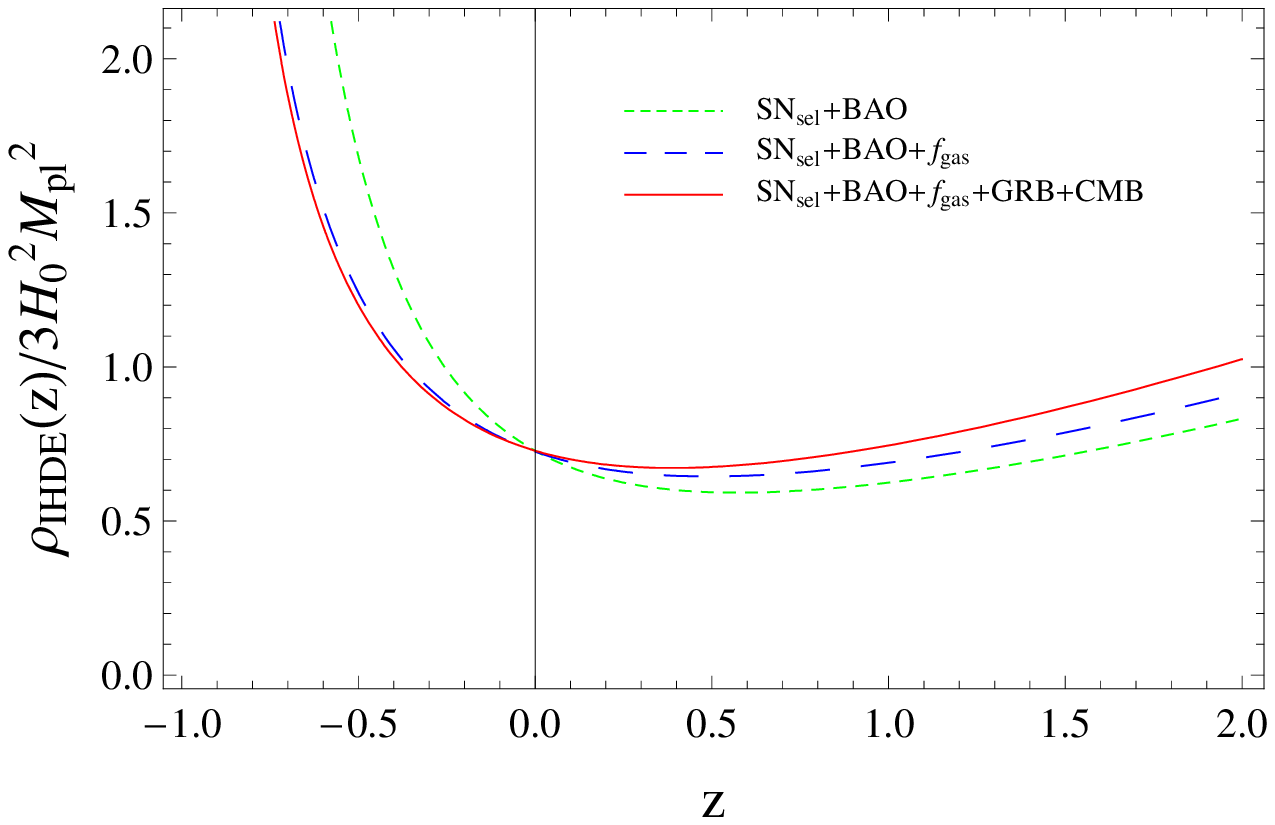}
}
\caption{Effective equation of state and density evolution for the best fit
IHDE models.}
\label{fig:IHDE}
\end{figure}

\begin{table}[h]
\caption{The Fitting results the IHDE model}%
\begin{ruledtabular}
\begin{tabular}[t]{c|c|c|c}
 & $\rm SN+BAO$ & $\rm SN+BAO+f_{gas}$ & $\rm SN+BAO+f_{gas}+GRB+CMB$\\
\hline
 $\Omega_{m0}$ & $0.272^{+0.023}_{-0.022}$ & $0.275^{+0.021}_{-0.021}$ & $0.281^{+0.017}_{-0.017}$ \\
c & $0.592^{+0.204}_{-0.113}$ & $0.667^{+0.321}_{-0.164}$ & $0.692^{+0.135}_{-0.107}$\\
 $\alpha$ & $-0.020^{+0.145}_{-0.174}$ & $0.068^{+0.093}_{-0.120}$ & $-0.006^{+0.021}_{-0.024}$ \\
$\Delta \ln \mathrm{BE}$& $~0.41 \pm 0.12~$ &$ ~0.70\pm0.18~$& $ ~0.75\pm0.18~$\\
\end{tabular}
\end{ruledtabular}
\end{table}
\begin{figure}[ht]
\begin{center}
\includegraphics[width=3.2in,height=3.2in]{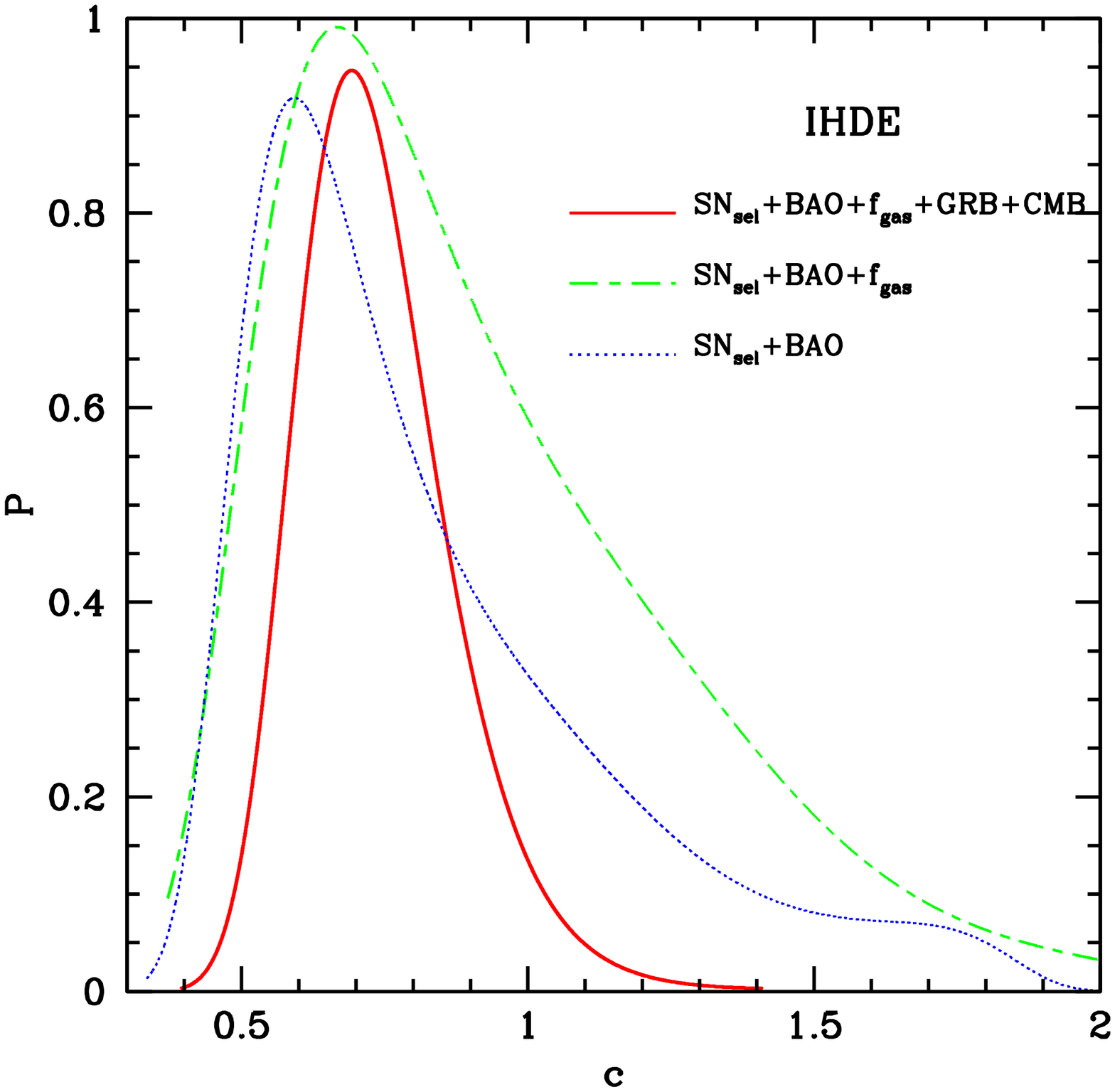} %
\includegraphics[width=3.2in,height=3.2in]{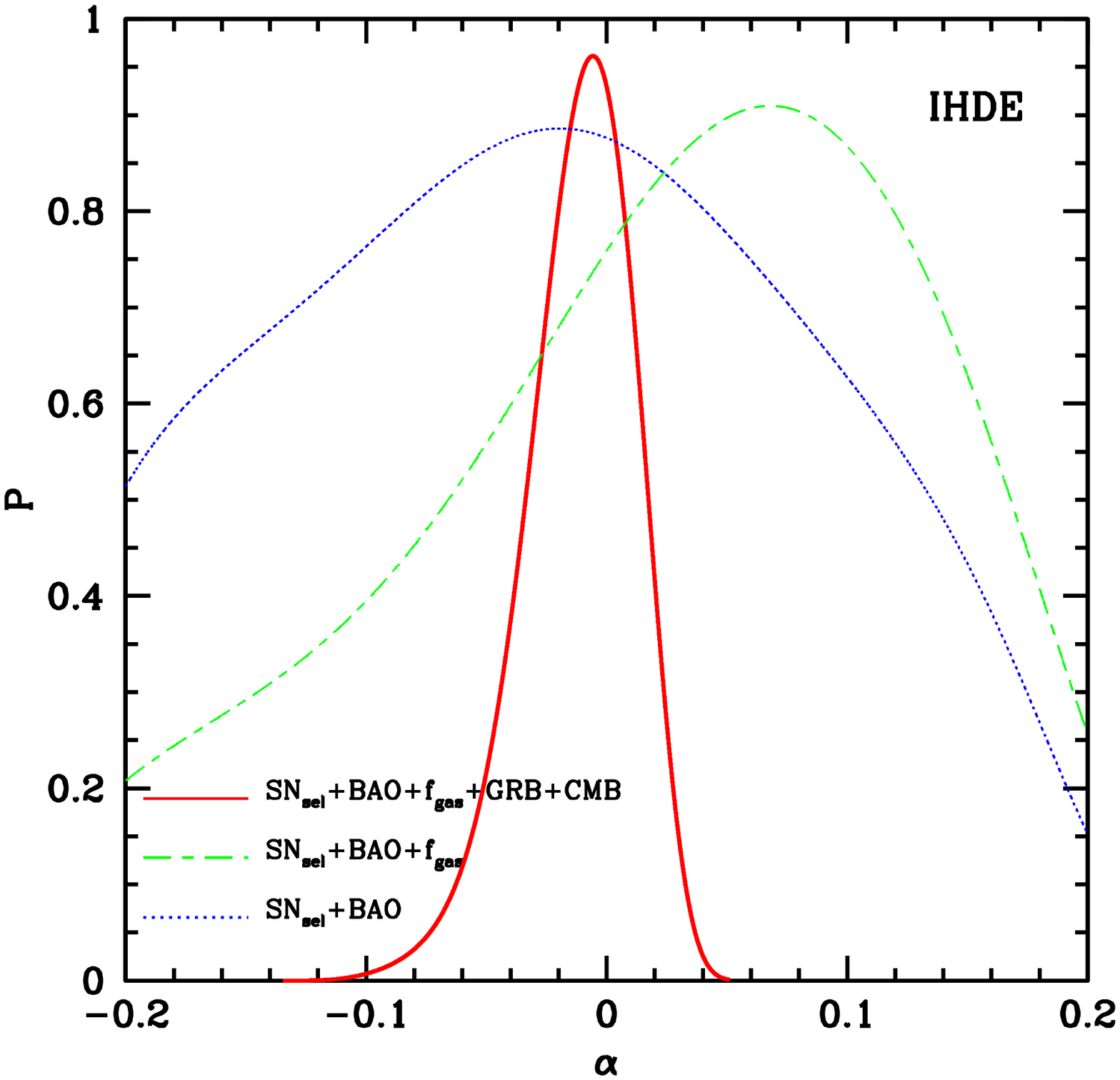}
\end{center}
\caption{The PDF of the parameters for the fits of IHDE model. Left: The PDF
for parameter $c$. Right: The PDF for parameter $\protect\alpha$. }
\label{fig:pdf_IHDE}
\end{figure}
\begin{figure}[ht]
\begin{center}
\includegraphics[height=3in]{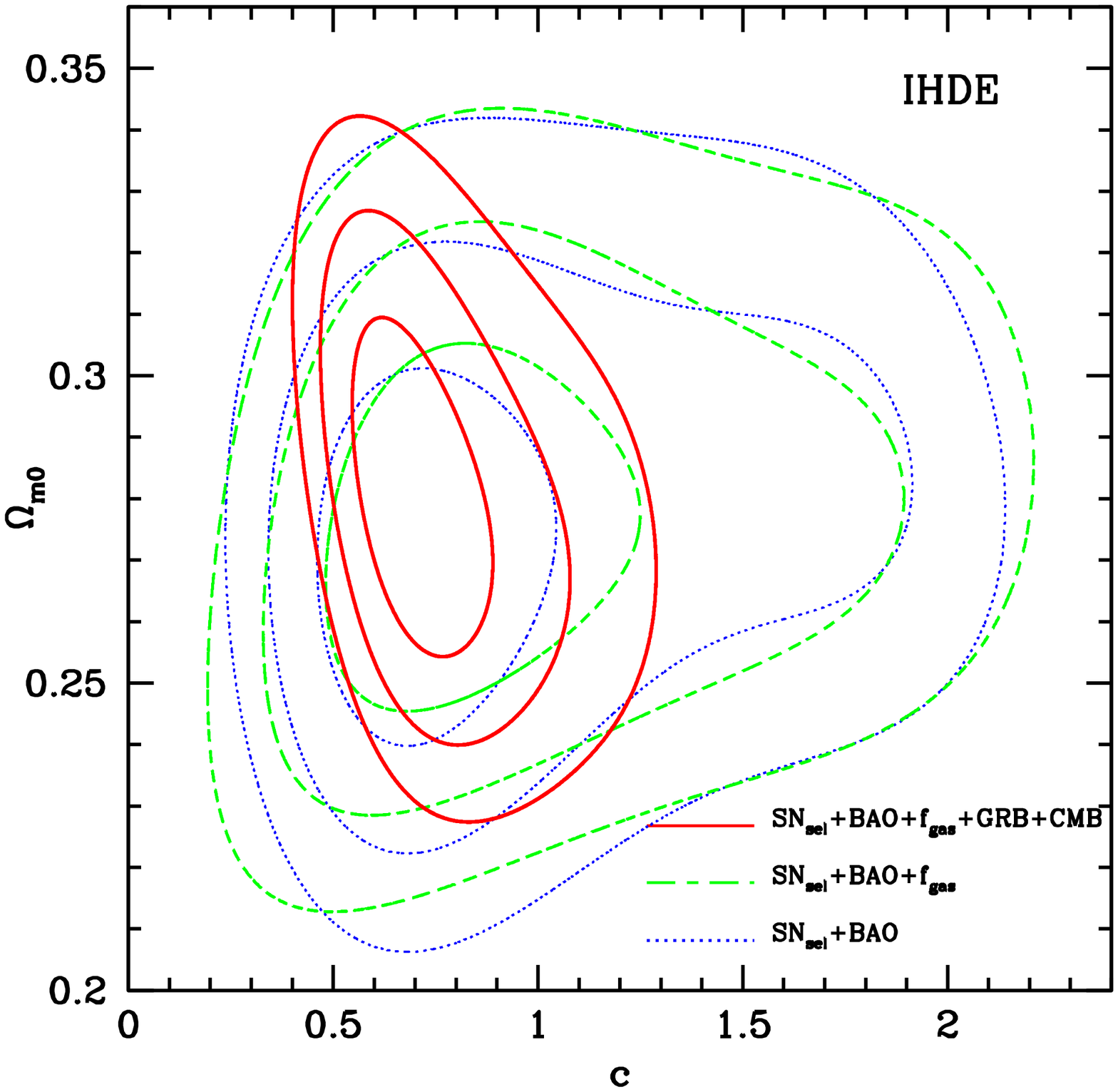}
\end{center}
\caption{The contour maps of $c$ vs. $\Omega _{m0}$ for IHDE with $1\protect%
\sigma (68.3\%)$, $2\protect\sigma (95.5\%)$ and $3\protect\sigma (99.7\%)$
confidence levels.}
\label{fig:m0contour}
\end{figure}
In Table II, we give the best fit value and the $1\sigma $ error of the IHDE
model parameters, as well as the value of $\ln \Delta \mathrm{BE}$ for the
three data set combinations. We plot the probability distribution function
(PDF) of the IHDE model in Fig.~\ref{fig:pdf_IHDE}. We also plot the $%
\Omega_{m0}-c$ contours in Fig.~\ref{fig:m0contour}. The evolution of the
effective equation of state and the relative density of the dark energy are
plotted in Fig.~\ref{fig:IHDE}.
Similar to the case of HDE models, the best fit of $\Omega _{m0}$ for the
IHDE models is around 0.27~0.28 for all three data set combinations, as can
be seen from Fig.~\ref{fig:m0contour}. However, for the different data set
combinations, the distribution of $c$ and $\alpha$ are fairly different (see
Fig.~\ref{fig:pdf_IHDE} and Table II). The peaks of the PDF for $\alpha$ are
different for the three data set combinations. Furthermore, the evolution of
the equation of state for the three data set combinations are also very
different (see Fig.~\ref{fig:IHDE}).
\begin{figure}[tbp]
\begin{center}
\includegraphics[width=4.2in,height=2.7in]{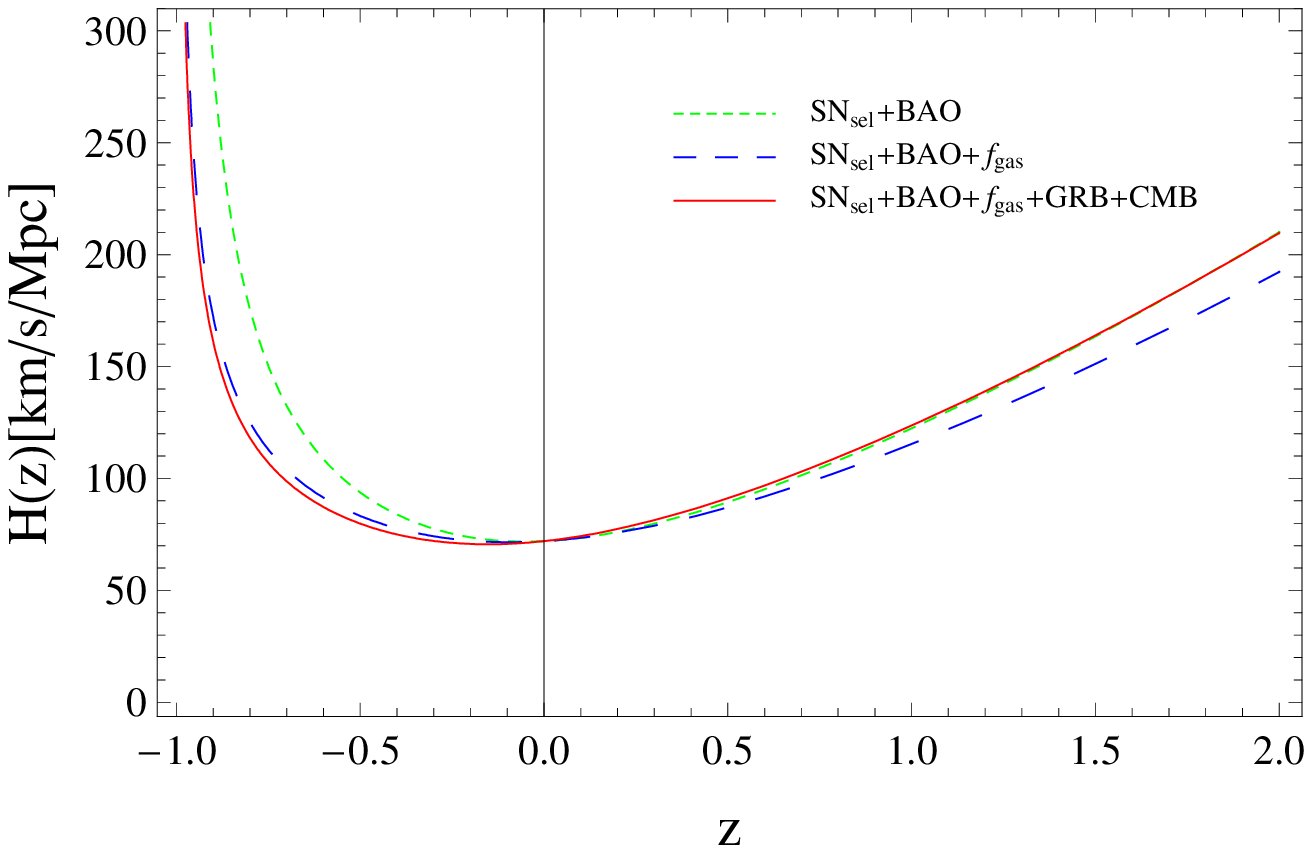}
\end{center}
\caption{The expansion rate $H(z)$ for the best fit parameters of the IHDE
model.}
\label{fig:H_IHDE}
\end{figure}

\begin{figure}[hb]
\includegraphics[scale = 0.4]{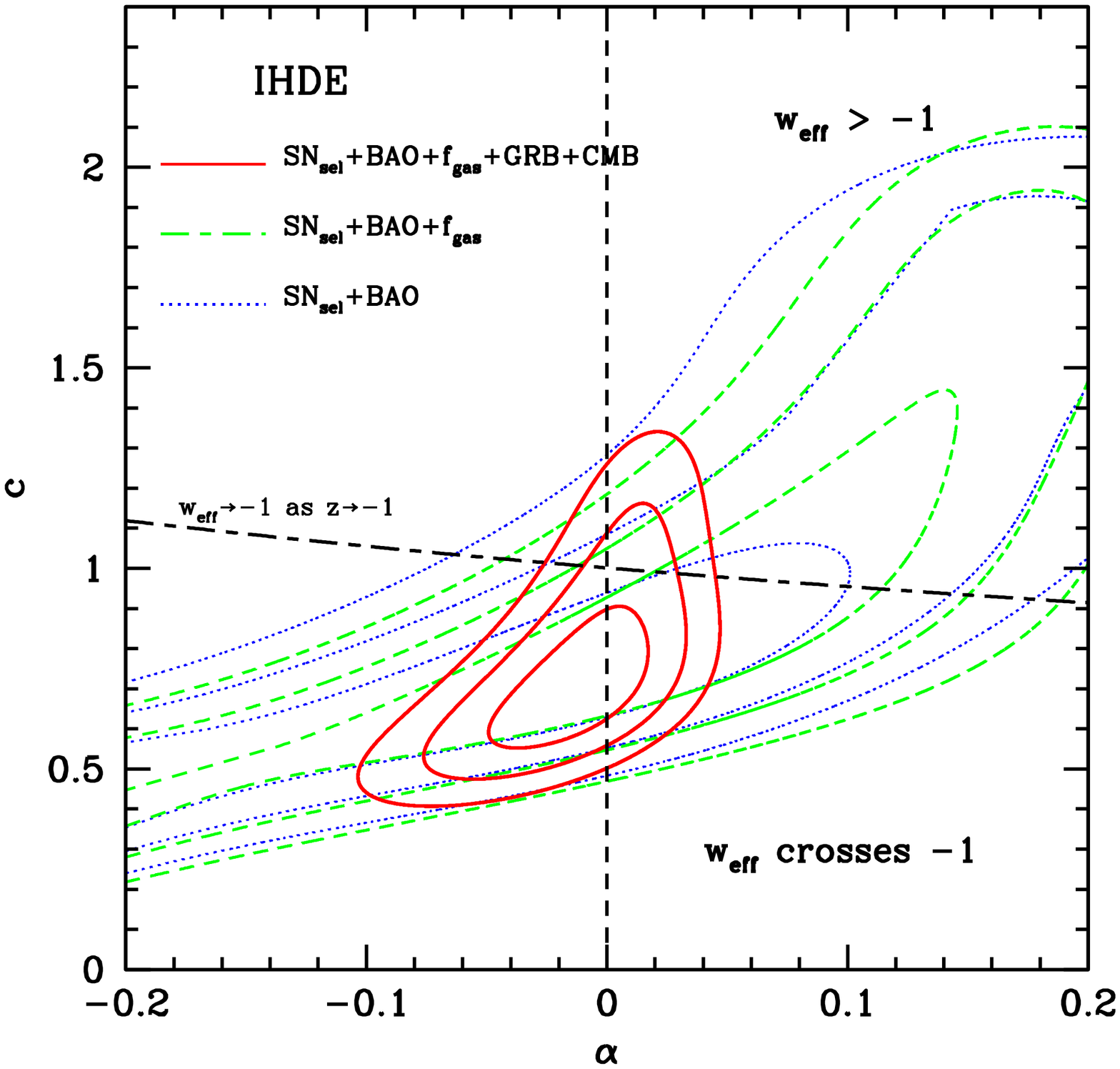}
\caption{The contour maps of $\protect\alpha $ vs.c for IHDE with $1\protect%
\sigma (68.3\%)$, $2\protect\sigma (95.5\%)$ and $3\protect\sigma (99.7\%)$
confidence levels. The black dot-dashed curve denotes $w_{\mathrm{eff}}=-1$
when $z\rightarrow -1$ with $\Omega _{\mathrm{IHDE0}}=0.73$, and the region
below (over) it means $w_{\mathrm{eff}}$ will (not) cross -1 during finite
time.}
\label{fig:alpha_c_i}
\end{figure}
Does this mean that the three different data sets indicate very different
expansion behavior? If so, this would indicate that the three data sets may
be inconsistent with each other. However, we have seen that for the case of
HDE the three different data set combinations yield similar fitting
parameters, indicating that they are basically consistent with each other.

To understand the origin of this difference, we plot the expansion rate in
Fig.~\ref{fig:H_IHDE} for the best fit model parameters of the three data
set combinations. While there are differences, one can see qualitatively the
three set of curves are similar to each other. This shows that the different
data sets are not inconsistent.

The reason of the difference seems to be due to parameter degeneracy in the
IHDE model. In this model, we have two parameters $c$ and $\alpha$, it
appears that different combination of these two parameters may lead to
similar dynamical behavior. At first sight, this seems to be in disagreement
with our analysis in \S 2, where the effective equation of state $w_{eff}$
of the dark energy was given in terms of these two parameters. However, the
equation of state $w_{eff}$ of the dark energy is not the only thing
affecting the expansion. In this model, the interaction between dark matter
and dark energy induces a non-zero effective equation of state for the dark
matter. Thus, although the effective equation of state for the dark energy
looks very different, the change in the dark matter equation of state
compensates part of this difference. Further, this model still has some
shortcomings since $\Omega _{de}$ could quantitatively evolve to region
larger than $1$ if $\alpha <0$, i.e. if dark matter decays into dark energy,
because the interaction term does not concern the energy density of dust
matter \cite{Miao08}. At this case, the IHDE model is only effective in the
period prior to the Big Rip. We will improve our work and discuss the
possible interaction forms in the next several papers.

The contour map for $c$ and $\alpha$ is plotted in Fig.~\ref{fig:alpha_c_i}.
From the Fig, the contours corresponding to the data sets SN+BAO+${f_{gas}}$%
+GRB+CMB are much tighter than the other data sets. This is because we use
the data set GRB and CMB, which has very large redshift distribution so they
break the degeneracy of the parameter $c$ and $\alpha$. We also mark the $%
\alpha-c$ values for which $w_{eff}=-1$ as a dashed line. This forms the
dividing line between quintessence-like and phantom-like behavior. For the
best fit parameters of all three data set combinations, $c \sim 0.6 < 1$.
The value of $\alpha$ varies more, but all consistent with being 0 within $%
1.5\sigma$, there is no strong evidence for the presence of interaction. For
the SN+BAO+$f_{gas}$+GRB+CMB, the PDF of parameter $\alpha$ and the best fit
values strongly suggest the evidence for the interaction is very weak. In
any case, for all three data set combinations, the best fit value resides in
the phantom-like region, although a large area of quintessence-like region
is also allowed.

The IHDE model is mildly favored over the $\Lambda$CDM model according to
the BE criterion, the evidence is slightly stronger than the HDE model case,
but not yet sufficient for drawing strong conclusions.

\section{Conclusion}

In this paper we firstly gave a brief review of holographic dark energy
model for both the non-interacting case and interacting case. We introduced
a new interacting term $Q=3\alpha H\rho _{\mathrm{de}}$ and the
non-interacting case could be viewed as the special case with $\alpha =0$.
We derived the equations for the evolution of $\Omega _{de}$ and $H(z)$, and
illustrated the dynamical behavior of these models by chosen some
representative values of the parameters $c$ and $\alpha$. The condition for
the model to have ``Big Rip'' is determined.

Secondly, we utilize several data sets from the resent observations to
constrain the models. Our data sets consist of the selected 182 high-quality
type Ia supernovae, the baryon acoustic oscillation measurement from the
Sloan Digital Sky Survey, the latest X-ray gas mass fraction data from $%
Chandra$ observations, 27 GRB samples generated with $E_{peak}-E_{\gamma }$
correlation, and the CMB shift parameter from WMAP three years result. We
used the MCMC technique to simulate the posterior probability of the model
parameters. The best-fits for the three data sets are given in Table 1 for
the HDE model, and Table 2 for the IHDE model. We also give the probability
distribution of the parameters and the contour maps for the HDE and IHDE
models.

Next, we utilize the Bayesian evidence (BE) as a model selection criterion
to compare the holographic models with $\Lambda $CDM model for the three
data sets. The BE is particularly appropriate for comparing models with
different number of parameters. Both the HDE and the IHDE model are mildly
favored by the current observational data set, although the evidence is
weak. For both the HDE and IHDE models, the data favors ``quintom'' behavior
slightly, i.e. the dark energy initially has $w_{eff}>-1$, but eventually
crossing the phantom dividing line, and the model ends with a ``Big Rip''.
However, quintessence-like behavior is also still allowed with the present
data.

In brief, we conclude that according to the combined measurements data, the
holographic dark energy model, especially the interacting holographic dark
energy model is mildly favored by the observations, and for the best fit
model the equation of state for both the HDE and IHDE crosses $-1$, for
which the Universe ends up in a Big Rip.

\section*{Acknowledgements}

We would like to thank Bin Wang, Chunshan Lin, Feng-Quan Wu, Hao Wei, Miao
Li, Rong-Gen Cai, Steven Allen, Xin Zhang, Yi Wang and Yungui Gong for
helpful discussions. One of the authors (Yin-Zhe Ma) also thanks Hao Yin,
Jian Ma and Nan Zhao for the help of computer program. Yin-Zhe Ma and Xuelei
Chen thank "UCLA 2008 dark matter and dark energy" advisory committee at Los
Angeles for hospitality during their stay when they present their work on
the symposium. Our MCMC chain computation was performed on the
Supercomputing Center of the Chinese Academy of Sciences and the Shanghai
Supercomputing Center. This work is supported by the National Science
Foundation of China under grants 10525314, 10325525, 90403029 and 10525060,
the Key Project Grant 10533010, by the Chinese Academy of Sciences under
grant KJCX3-SYW-N2, and by the Ministry of Science and Technology under the
national basic sciences program (973) under grant 2007CB815401.

\clearpage

\end{document}